\documentclass[letterpaper, oldversion]{aa}  

\usepackage{aalongtable}
\usepackage{natbib}\usepackage{graphicx}      
\bibpunct{(}{)}{;}{a}{}{,}                

\newcommand{\kms}{km\,s$^{-1}$}     
\newcommand{\sqcm}{cm$^{-2}$}  
\newcommand{\lya}{Lyman-$\alpha$}
\newcommand{\hi}{\ion{H}{i}} 
\newcommand{\hw}{\ion{H}{ii}} 
\newcommand{\os}{\ion{O}{vi}} 
\newcommand{\cf}{\ion{C}{iv}}
\newcommand{\sif}{\ion{Si}{iv}}
\newcommand{\nf}{\ion{N}{v}}
\newcommand{\none}{\ion{N}{i}}
 
\newcommand{\ntot}{164} 
\newcommand{\ndla}{141} 
\newcommand{\nsub}{23}  
\newcommand{\ndet}{fourteen} 
\newcommand{\ncomp}{39} 
\newcommand{\zabs}{$z_{\rm abs}$}
\newcommand{\zqso}{$z_{\rm qso}$}

\begin{document}

\title{Metal-enriched plasma in protogalactic halos\thanks{Based 
  on observations taken with the Ultraviolet and Visual Echelle
  Spectrograph (UVES) on the Very Large Telescope (VLT) Unit 2
  (Kueyen) at Paranal, Chile, operated by the European Southern
  Observatory (ESO), and with the 
  High Resolution Echelle Spectrograph (HIRES) and Echelle
  Spectrograph and Imager (ESI) instruments located at the 
  W. M. Keck Observatory on Mauna Kea, Hawaii.}}
\subtitle{A survey of \nf\ absorption in high-$z$ damped \&
  sub-damped \lya\ systems}
\author{Andrew J. Fox\inst{1}, Jason X. Prochaska\inst{2}, 
  C\'edric Ledoux\inst{1}, Patrick Petitjean\inst{3}, 
  Arthur M. Wolfe\inst{4}, \& Raghunathan Srianand\inst{5}}
\institute{
  European Southern Observatory, Alonso de C\'ordova 3107, Casilla
  19001, Vitacura, Santiago, Chile; afox@eso.org \and 
  University of California/Lick Observatory, UC Santa Cruz, 1156 High
  Street, CA 95064 \and    
  Institut d'Astrophysique de Paris, UMR7095 CNRS,
  UPMC, 98bis Blvd Arago, 75014 Paris, France \and
  Center for Astrophysics and Space Sciences, UC San Diego, 9500
  Gilman Drive, CA 92093 \and  
  Inter-University Centre for Astronomy and Astrophysics, Post Bag 4,
  Ganesh Khind, Pune 411 007, India}  

\date{Received 23 January, 2009 / Accepted 7 May, 2009}  

\authorrunning{Fox et al.}
\titlerunning{\nf\ in DLAs and sub-DLAs} 

\abstract{
We continue our recent work to characterize the plasma content of
high-redshift damped and sub-damped \lya\ systems (DLAs/sub-DLAs), 
which represent multi-phase gaseous (proto)galactic disks and halos
seen toward a background source. 
We survey \nf\ absorption in a sample of 91 DLAs and 18 sub-DLAs in 
the redshift range 1.67$<$\zabs$<$4.28 with unblended coverage of the
\nf\ doublet. Our dataset includes high-resolution (6--8~\kms\ FWHM)
quasar spectra obtained with VLT/UVES and Keck/HIRES, together with
medium-resolution ($\approx$40~\kms\ FWHM) quasar spectra from Keck/ESI. 
In DLAs, we find eight secure \nf\ detections,
four marginal detections, and 79 non-detections, for 
which we place 3$\sigma$ upper limits on the \nf\ column density.
The detection rate of \nf\ in DLAs is therefore 13$^{+5}_{-4}$\%. 
Two sub-DLA \nf\ detections are found among a sample of 18, at a
similar detection rate of 11$^{+15}_{-7}$\%. 
We show that the \nf\ detection rate is a strong function of 
neutral-phase nitrogen abundance, increasing by a factor of $\approx$4 at
[N/H]=[\none/\hi]$>\!-2.3$. 
The \nf\ and \cf\ component $b$-value distributions in DLAs are
statistically similar, but the median $b$(\nf) of 18~\kms\
is narrower than the median $b$(\os) of $\approx$25~\kms.
Some $\approx$20\% of the \nf\ components have $b\!<\!10$~\kms\ and
thus arise in warm, photoionized plasma at log\,(T/K)$<$4.92; local
sources of ionizing radiation (as opposed to the extragalactic
background) are required to keep the cloud sizes physically
reasonable. The nature of the remaining $\approx$80\% of (broad) \nf\
components is unclear; models of radiatively-cooling
collisionally-ionized plasma at log\,($T$/K)=5.2--5.4 are fairly
successful in reproducing the observed integrated high-ion column
density ratios and the component line widths, but we cannot rule out
photoionization by local sources.
Finally, we identify several unusual DLAs with extremely low 
metallicity ($<$0.01 solar) but strong high-ion absorption
[log\,$N$(\nf)$>$14 or log\,$N$(\os)$>$14.2],
which present challenges to either galactic inflow or outflow models.}
\maketitle

\section{Introduction}
Ever since the epoch of reionization, the vast majority of all
baryons in the Universe have existed in either warm or hot plasma.
At high redshift ($z\!\sim\!3$), the baryon budget is dominated by
plasma in the warm, photoionized intergalactic medium 
\citep[IGM, traced by the \lya\ forest;][]{Pj93, Sh94, We97, Ra97}.
By the time the Universe has evolved to the present day,
structure formation and feedback from star formation have
redistributed the baryons; simulations find they are fairly evenly
shared between condensed gas in galactic structures, warm 
diffuse plasma, and warm-hot shock-heated plasma \citep[][]{CO99,
CO06, Da01}. Although questions remain about whether the detailed
baryon budget has been confirmed observationally \citep{Fu98, Br07,
PT08}, the finding that most baryons exist in plasma is robust.
Therefore, characterizing the plasma phases observationally is
crucial to understand the history of 
baryonic matter.

Counter-intuitively, a good place to find astrophysical plasma is in
those potential wells where large quantities of \emph{neutral} gas
reside. Such locations can be identified as damped \lya\ (DLA)
systems, the well-studied category of quasar absorbers defined by a
neutral hydrogen column density log\,$N$(\hi)$>$20.3 \citep[see review
by][]{Wo05}. There are several lines of evidence suggesting that DLAs
represent star-forming galaxies (or their halos) seen in absorption,
including the identification of the host galaxy in many low-$z$
\citep{lB97, Ra03, CL03, Ch05} and high-$z$ \citep{Wa01, Mo02, Mo04} DLAs,
the presence of metals \citep[e.g.][]{Pr03a, Pr03b} and molecular gas
\citep[e.g.][]{Le03, Sr05, No08} in DLAs, and a significant cross-correlation
signal with Lyman Break Galaxies at $z\!\approx\!3$ \citep{Co06a,
Co06b}. Absorbers with slightly lower \hi\ columns
[19.0$<$log\,$N$(\hi)$<$20.3] are referred to as sub-DLAs 
\citep[or Super Lyman Limit systems;][]{DZ03, Px03, Px05, OM07}. 
Sub-DLAs are also thought to be arise in galaxies (or galaxy halos),
but are generally treated as a separate class of objects than
DLAs, because of a partly ionized (rather than neutral) \hi\ phase
\citep{PW96, Px07}.

The presence of plasma in DLA and sub-DLA host galaxies is
indicated by the detection of \cf\ and \sif\ absorption at $z_{\rm DLA}$
in {\it all} DLAs and sub-DLAs with \cf\ spectral coverage
\citep[][hereafter Paper I and Paper II]{Lu96, PW96, PW99, Pr99, Le98,
  WP00a, Pr02a, DZ03, Px03, Ri05, Le08, Qu08, Fo07a, Fo07b}.  
The \cf\ and \sif\ kinematic structure
is almost always distinct from that of the low ions \citep{WP00b},
indicating the high ions trace different regions. 
Recently, a new plasma phase was revealed by detections of
\os\ absorption in thirteen DLAs \citep[Paper I;][]{Le08} and three sub-DLAs
\citep{Si02, Fo07c} at $z\!\approx\!2$--3. 
Since none of the robustly detected \os\ components in DLA/sub-DLAs
are narrow (all have $b\!>\!10$~\kms), and that photoionization models
fail to simultaneously explain the \os\ and other high-ion column
densities \citep[Paper I;][]{Le08}, 
the \os\ detections constitute the discovery
of warm-hot, collisionally-ionized plasma in the DLA host
galaxies\footnote{The standard definition of warm-hot is
  $T$=10$^5$--10$^7$K; however, \os\ is only produced by  collisions
  at the lower end of this range, at $T\!\sim\!10^5$--10$^6$K.}. 

Since cosmological simulations predict that little warm-hot plasma
has formed at $z\!\ge\!3$ \citep{Da01}, its presence at this epoch
raises some interesting questions. Does warm-hot plasma in galaxy
halos form by feedback processes from star formation and supernovae
\citep{OD06, KR07, Fa07}, rather than through the shock-heating of
infalling material \citep[hot-mode accretion;][]{BD03, Ke05, DB06}? 
Does warm-hot plasma in galaxy halos contain a significant fraction of
the high-redshift metal budget \citep{Fe05, SF08}?
How do those halos evolve with time?
Continued observational studies of high-ion absorption lines in galactic
structures at low- and high-$z$, together with techniques to
distinguish between warm photoionized- and hot collisionally-ionized
absorbers, are needed to answer these questions.  

After O$^{+5}$, the second-most-highly-ionized species accessible in
the rest-frame UV is N$^{+4}$, whose resonance doublet is \nf\
$\lambda\lambda$1238.821, 1242.804. Though only 77.5\,eV is required
to create N$^{+4}$ versus 113.9\,eV to create O$^{+5}$, it is
difficult to photoionize \nf\ with starlight, since the flux of
stellar photons falls rapidly above 54~eV (228~\AA), the \ion{He}{ii}
ionization edge \citep{BH86}. Thus interstellar \nf\ absorption traces
plasma that is either hot or subject to a hard
photoionizing spectrum. The \nf\ doublet has the significant advantage
of lying less deep into the \lya\ forest than \os, so the signal-to-noise is
higher and the level of blending is lower. However, it is
well-documented that nitrogen tends to be under-abundant in DLAs 
\citep{Pe95, Pe02a, Pe08, Ce98, Ce03, Lu98, Pr02b, HP07, Pj08} and in 
low-metallicity environments in general. The low nitrogen abundance
coupled with the intrinsic weakness of both lines of the \nf\ doublet 
makes detecting \nf\ in DLAs challenging, but not impossible:
\nf\ detections have been reported in two intervening DLAs (Paper I) and
three highly proximate DLAs \citep[within $\sim$100~\kms\ of, or
beyond, $z_{\rm QSO}$; Paper I; ][]{Rx07, He09}.

The goal of this paper is to survey \nf\ in DLAs and
  sub-DLAs\footnote{When we refer to \nf\ (or any high ion) \emph{in}
  a DLA, we are not implying that the \nf\ and \hi\ lines form in the
  same regions of gas. Rather, ``\nf\ in DLAs'' is a convenient
  shorthand for ``\nf\ in the multi-phase structures whose neutral
  phase is seen in \lya'', or equivalently ``\nf\ in DLA host galaxies''.}.
We structure the paper as follows. In \S2 we describe the three datasets
combined to produce our sample, together
with our data handling and measurement techniques. In \S3 we summarize
the statistics on the frequency of detection of \nf. In \S4 we
describe the measurement of the nitrogen abundance in cases where the \none\
lines are available. In \S5 we present
our results, investigating how the \nf\ detections and non-detections
depend on other observational properties of DLAs. We discuss the
implications of our results in \S6, and summarize our principal
findings in \S7. Throughout this paper, we use the standard notation
[X/Y]$\equiv$log\,[$N$(X)/$N$(Y)]--log\,[$N$(X)/$N$(Y)]$_\odot$ when
discussing 
elemental abundances. We use the photospheric solar abundance table
compiled by \citet{Lo03} unless otherwise noted.

\section{Observations and Data Handling}
\subsection{Origin of Data}
To create our sample, we combined pre-existing datasets from the
following three instruments: \\
(1) The Ultraviolet/Visual Echelle Spectrograph \citep[UVES;][]{De00}
mounted on the 8.2\,m Very Large Telescope Unit 2 (Kueyen) on
Cerro Paranal, Chile.
Our UVES sample of DLAs and sub-DLAs has been built over the
last few years for the purpose of studying elemental abundances and
molecular gas, and is described in \citet{Le03, Le06}, 
supplemented with several spectra from the Hamburg-ESO survey 
\citep{Sm05}. One additional sub-DLA was taken from
the sample of \citet{Be04}. The data reduction was conducted using a
customized version of the MIDAS pipeline written by \citet{Ba00}. The
UVES data have a velocity resolution of 6.6~\kms\ (FWHM), and  
a mean pixel size of 2.1~\kms\ (after using the 2x2 binning mode). 
UVES has excellent response in the blue, where \nf\ absorption at
$z$=2--3 is observed.\\  
(2) The High Resolution Echelle Spectrograph \citep[HIRES;][]{Vo94}
located on the Keck I telescope on Mauna Kea, Hawaii. Our HIRES
dataset is based on 
the sample published in \citet{Pr07}, with several new QSO spectra taken
from \citet{Wo08}. The data reduction was conducted using the MAKEE
package (developed and distributed by T. Barlow) for datasets taken with the
original Tektronix CCD, and using the HIRedux package \citep{Pr07} for data
obtained with the newer CCD mosaic.
The HIRES data have a velocity resolution of either 6.6 or
8.0~\kms\ (FWHM), depending on the choice of slit width, 
and are unbinned, with a mean pixel size of 1.8~\kms.\\ 
(3) The Echelle Spectrograph and Imager \citep[ESI;][]{Sh02} 
mounted on the Keck II telescope. These data, published
in \citet{Pr07}, were obtained using a 0.5\arcsec\ or
0.75\arcsec\ slit, resulting in a velocity FWHM of 33 or 44~\kms,
respectively, and are unbinned with a pixel size of 10.0~\kms. 
The ESI data were reduced with the ESIRedux package developed in IDL 
and described in \citet{Pr03a}.

\subsection{Formation of sample}
We began by selecting all DLAs and sub-DLAs from our sample with
spectra covering \nf, i.e. covering the wavelength
1238$\times$(1+$z_{\rm abs}$)~\AA, where $z_{\rm abs}$ is defined
by the wavelength of strongest absorption in the low-ionization metal lines. 
After removing duplicate systems present in both UVES and HIRES
samples (for which we generally took the UVES data, unless the HIRES S/N was
significantly higher), we ended up with \ndla\ individual 
DLAs and \nsub\ sub-DLAs; of these \ntot\ absorbers, the data are from
UVES in 112 cases, from HIRES in 35, and from ESI in 17. 
The larger number of DLAs than sub-DLAs simply reflects the way the samples
were constructed; it does not reflect their relative incidence of
occurrence along random sight-lines.
Continua were fitted in the vicinity of \nf\ (typically
using a spectral region covering several thousand \kms) using
low-order polynomials. We searched for the presence of \nf\
absorption in a velocity interval of $\pm$400~\kms\ around $z_{\rm abs}$,
and then classified each system into one of the following categories: \\

\noindent 1. Secure \nf\ detection (eight DLAs and two sub-DLAs), 
confirmed by consistent optical depth profiles in both members of the
doublet (\nf\ $\lambda\lambda$1238.821, 1242.804). \\ 
2. Marginal \nf\ detection (four DLAs), seen in one line of the 
doublet only (the other being blended) but with a profile consistent
with the \cf\ absorption seen at that redshift.\\ 
3. No detection (79 DLAs and 16 sub-DLAs); 
3$\sigma$ upper limit on column density measured.\\ 
4. Completely blended (50 DLAs and 5 sub-DLAs); both doublet lines 
contaminated,  with no useful information on the presence or absence
of \nf.\\  

The systems in category 4 were discarded. 
Note that a system can enter category 1 even if the \nf\ data are
partly blended. This occurs when \nf\ absorption is clearly
identified in some portion of the line profiles but is blended at
other velocities. In these cases the reported \nf\ column densities
are formally lower limits, since additional absorption could be hidden
in the blended regions.
Examples of each category are given in Figure 1.
We note that the DLA at $z_{\rm abs}$=1.82452 toward \object{Q1242+0006}
shows \nf\ in two components at $-$709$\pm$2~\kms\ and $-$842$\pm$2~\kms,
each with coincident \cf\ absorption. This absorber was
included in our sample despite its large velocity separation from the
neutral gas, since there is a good probability it is physically
connected to the DLA: there are several examples in the literature of
DLAs with \cf\ absorption extending over 500--1\,000~\kms\ 
around the DLA redshift \citep[][Paper II]{PW99, WP00a, Pj02}.

\begin{figure*}[!ht]
\includegraphics[width=18.5cm]{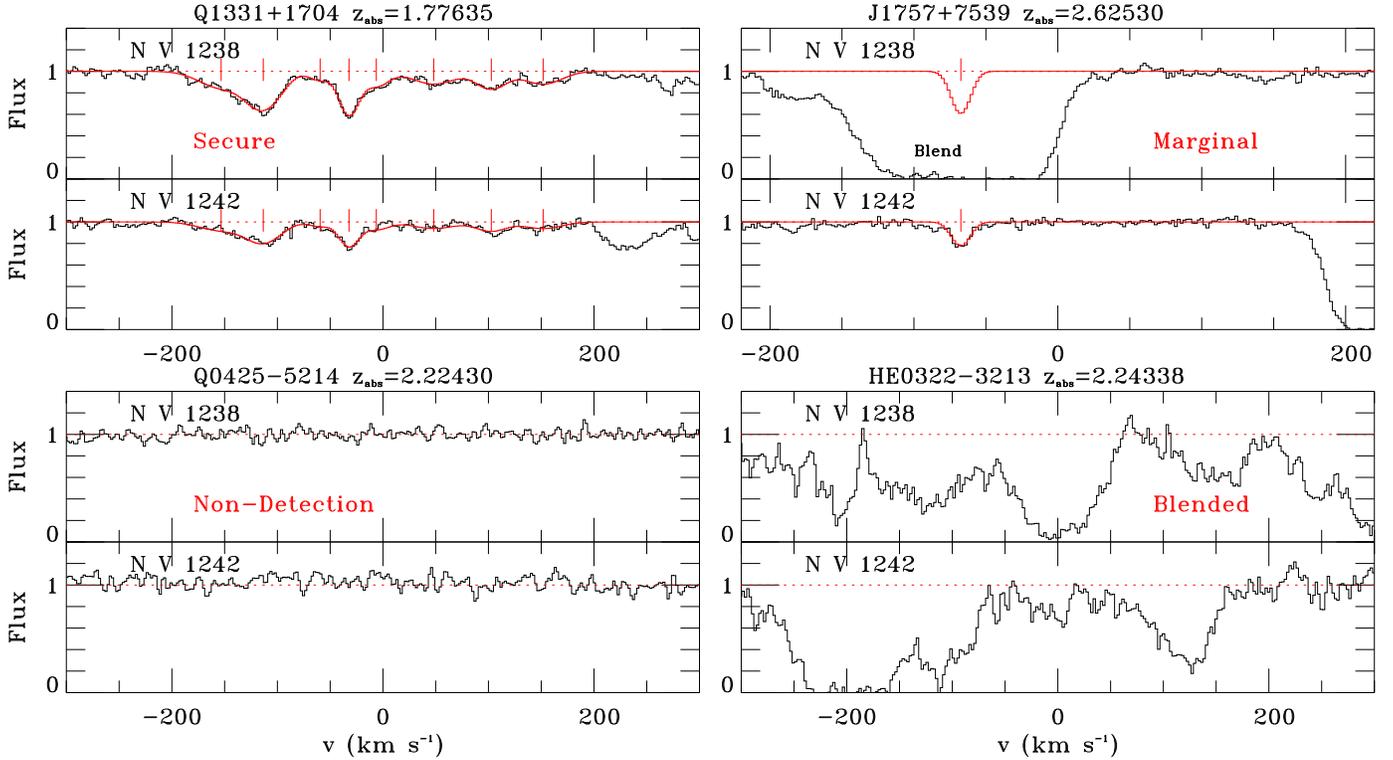}
\caption{Example DLA \nf\ normalized absorption-line spectra for our
  four categories: secure detection, marginal detection (one doublet
  line only), non-detection, and blended. For the detections, best-fit
  Voigt profile models are shown in red, with component centers marked
  with ticks. Marginal \nf\ detections coincide in velocity with
  components in \sif\ and \cf\ (whereas blended \nf\ profiles do not). See the
  Appendix for the full absorption-line spectra.} 
\end{figure*}

\subsection{Measurements}
\begin{table}
\begin{minipage}[t]{8.5cm}
\caption{Voigt component fits to DLA/sub-DLA \nf}
\centering
\begin{tabular}{lc ccc}
\hline \hline
QSO & $z_{\rm abs}$\footnote{$\dagger$ denotes a proximate DLA ($<$5\,000~\kms\ from QSO).} & $v_0$\footnote{Velocity centroid relative to $z_{\rm abs}$.} & $b$ & log\,$N$\footnote{$^*$ denotes a marginal \nf\ detection.}\\
    &               & (\kms) & (\kms) & ($N$ in \sqcm)\\
\hline
{\bf DLAs}\\
     J1014+4300 &               2.95880 & $-$ 13$\pm$2 &  29$\pm$3 & 13.20$\pm$0.04$^*$ \\
     J1211+0422 &               2.37655 & $-$ 98$\pm$12 &  21$\pm$10 & 13.12$\pm$0.17     \\
                &                       & $-$ 67$\pm$16 &  20$\pm$14 & 12.92$\pm$0.17     \\
                &                       & $-$ 14$\pm$2 &  18$\pm$2 & 13.33$\pm$0.03     \\
     J1757+7539 &               2.62530 & $-$ 67$\pm$2 &   7$\pm$1 & 13.20$\pm$0.03$^*$ \\
    J2100--0641 &      3.09240$\dagger$ &      6$\pm$7 &  74$\pm$10 & 13.62$\pm$0.05$^*$ \\
    Q0042--2930 &               1.80947 & $-$ 89$\pm$2 &   6$\pm$1 & 13.59$\pm$0.04$^*$ \\
    Q0450--1310 &               2.06655 & $-$  3$\pm$2 &  17$\pm$1 & 13.29$\pm$0.02     \\
                &                       &     30$\pm$2 &   9$\pm$1 & 12.82$\pm$0.04     \\
                &                       &     88$\pm$2 &  22$\pm$3 & 12.99$\pm$0.05     \\
    Q0528--2505 &      2.81115$\dagger$ & $-$ 41$\pm$5 &  51$\pm$4 & 13.43$\pm$0.06     \\
                &                       &     12$\pm$2 &  18$\pm$1 & 13.70$\pm$0.05     \\
                &                       &     27$\pm$2 &   8$\pm$2 & 12.87$\pm$0.13     \\
                &                       &     43$\pm$13 &  47$\pm$10 & 13.26$\pm$0.17     \\
    Q0551--3637 &               1.96221 & $-$111$\pm$2 &  13$\pm$3 & 13.26$\pm$0.11     \\
                &                       & $-$ 97$\pm$7 &  74$\pm$8 & 14.09$\pm$0.07     \\
                &                       & $-$ 61$\pm$6 &  21$\pm$14 & 12.84$\pm$0.47$^*$ \\
                &                       &      3$\pm$20 &  79$\pm$15 & 13.60$\pm$0.18$^*$ \\
     Q1242+0006 &               1.82452 & $-$842$\pm$2 &  11$\pm$1 & 13.90$\pm$0.02$^*$ \\
                &                       & $-$709$\pm$2 &  13$\pm$2 & 13.50$\pm$0.05     \\
     Q1331+1704 &               1.77635 & $-$153$\pm$11 &  28$\pm$10 & 13.21$\pm$0.22     \\
                &                       & $-$113$\pm$3 &  22$\pm$3 & 13.52$\pm$0.11     \\
                &                       & $-$ 59$\pm$5 &  12$\pm$8 & 12.61$\pm$0.25     \\
                &                       & $-$ 32$\pm$2 &  11$\pm$3 & 13.35$\pm$0.10     \\
                &                       & $-$  6$\pm$6 &  16$\pm$9 & 12.93$\pm$0.25     \\
                &                       &     48$\pm$4 &  27$\pm$9 & 13.08$\pm$0.11     \\
                &                       &    103$\pm$3 &  19$\pm$5 & 13.08$\pm$0.10     \\
                &                       &    152$\pm$4 &  24$\pm$6 & 13.03$\pm$0.09     \\
    Q2243--6031 &               2.33061 & $-$356$\pm$2 &  13$\pm$1 & 13.41$\pm$0.01     \\
                &                       & $-$320$\pm$2 &   9$\pm$2 & 12.87$\pm$0.04$^*$ \\
    Q2348--1444 &               2.27940 & $-$128$\pm$2 &   8$\pm$1 & 13.53$\pm$0.05     \\
                &                       & $-$ 98$\pm$2 &  18$\pm$1 & 14.29$\pm$0.01     \\
                &                       & $-$ 62$\pm$2 &   9$\pm$1 & 13.57$\pm$0.03     \\
                &                       &      2$\pm$2 &  13$\pm$2 & 13.36$\pm$0.04     \\
                &                       &     48$\pm$2 &  10$\pm$1 & 13.18$\pm$0.03     \\
                &                       &    137$\pm$9 &  19$\pm$3 & 13.96$\pm$0.10     \\
                &                       &    160$\pm$18 &  19$\pm$15 & 13.13$\pm$0.10     \\
\hline
{\bf Sub-DLAs}\\
    Q0237--2321 &               1.67234 & $-$ 37$\pm$2 &  12$\pm$1 & 13.33$\pm$0.02     \\
                &                       &     21$\pm$3 &  25$\pm$3 & 13.41$\pm$0.06     \\
                &                       &     54$\pm$4 &  14$\pm$9 & 12.62$\pm$0.15     \\
                &                       &    163$\pm$10 &  19$\pm$15 & 13.23$\pm$0.20     \\
                &                       &    183$\pm$3 &   9$\pm$7 & 12.87$\pm$0.15     \\
    Q1037--2704 &               2.13906 & $-$308$\pm$2 &  22$\pm$1 & 13.60$\pm$0.01     \\
                &                       &      5$\pm$2 &  40$\pm$3 & 13.29$\pm$0.03     \\
\hline
\end{tabular}
\end{minipage}
\end{table}

For each system, we identify a velocity integration range $v_-$ to $v_+$
(defined relative to the DLA redshift $z_{\rm abs}$) over which
the rest-frame equivalent width of absorption is measured. 
For \nf\ detections, these limits are identified by eye. For the \nf\
non-detections, we used (where possible) the 
velocity range where \cf\ absorption is observed. If \cf\ data were
unavailable, or if the \nf\ profiles were blended in the \cf\ region, we
adopted a range $-$50 to 50~\kms. The equivalent width is first
measured per pixel, with an error estimate that includes contributions
from statistical noise and continuum placement. The total equivalent
width is then formed by summing over the pixels in the chosen velocity
range, with the error on the total equivalent width formed by adding
the individual pixel errors in quadrature \citep[see appendix in][]{SS92}. 
We include an additional error of 2~m\AA\ in each equivalent width 
measurement, reflecting the uncertainty in the choice of velocity limits.

For the \nf\ detections, we fit Voigt components to the absorption 
in all available high ions using the software package
VPFIT\footnote{Available at http://www.ast.cam.ac.uk/$\sim$rfc/vpfit.html.}.
VPFIT operates simultaneously on the two doublet lines and accounts 
for the effects of instrumental resolution. 
The number of \nf\ components to fit was usually self-evident from
inspection of the line profiles, but in all cases 
the other high-ion profiles were considered when deciding how
many components to fit in the \nf\ model.
However, the \nf, \cf, \os, and \sif\ fits were each conducted
independently, i.e. no component centroids or line widths were tied
from one ion to the next.
There is always the possibility of adding extra components, but our
approach was to use the minimum number necessary.
For the twelve DLAs and two sub-DLA with \nf\ 
detections, the detailed parameters of our best-fit \nf\ Voigt
profiles are given in Table 1. 
For each component, we added (in quadrature) extra terms to the errors
reported by VPFIT, of 2.0~\kms\ (velocity centroid errors),
1.0~\kms\ ($b$-value errors) and 0.02\,dex (column density errors), to
reflect the minimum plausible uncertainties.
The absorption-line spectra
(together with the best-fit VPFIT models) are shown in the Appendix as
Figures A.1, A.2, and A.3 for DLAs and Figure A.4 for sub-DLAs. We include
all available high-ion data in each stack, as well as the profile of
\ion{Si}{ii} $\lambda$1304, \ion{Si}{ii} $\lambda$1808, 
\ion{S}{ii} $\lambda$1253 or \ion{Fe}{ii} $\lambda$1608, to trace the
low-ionization gas. Some of these data have already been published
\citep[][Paper I, Paper II]{Le08}, but we redisplay them here for
completeness, and because the other high-ion profiles are crucial
for interpreting the \nf.

For the \nf\ non-detections, i.e. cases where there is no absorption
present at 3$\sigma$ significance in the chosen velocity range, we
determined the 3$\sigma$ upper allowed value of the rest-frame
equivalent width; 
for example, if a measurement found $W_{\lambda}$=0$\pm$3~m\AA, we
adopted $W_{\lambda}^{3\sigma{\rm lim}}$=9~m\,\AA. 
We then converted $W_{\lambda}^{3\sigma{\rm lim}}$ into a 3$\sigma$ limit
on the column density assuming a linear curve-of-growth, i.e.   
$N_{\rm \nf}^{3\sigma{\rm lim}}=1.13\times10^{17}W_{\lambda}^{3\sigma{\rm lim}}
/\lambda_0^2 f$, where $W_{\lambda}^{3\sigma{\rm lim}}$ is in m\AA,
$\lambda_0$ is in \AA, and $N_{\rm \nf}$ is in cm$^{-2}$. 
The optically-thin assumption is justified since if any absorption on
the non-linear parts of the curve-of-growth were present, this would
preclude a non-detection. 
We verified these limits by using an alternative method to derive the
limiting equivalent width, following the formula given in
\citet{Wa96}, which computes $W_{\lambda}^{3\sigma{\rm lim}}$ as a
function of S/N, pixel size, and assumed line width $b$=18~\kms. 
The two methods were found to given broadly similar results, with a
mean absolute difference between the calculated values of
log\,$N_{\rm \nf}^{3\sigma{\rm lim}}$ of 0.22\,dex. 
In the Appendix, we present a summary of \nf\
measurements for all detections and non-detections in DLAs (Table A.1)
and sub-DLAs (Table A.2). In each system we present the result from
whichever of $\lambda$1238 or $\lambda$1242 gave the stronger constraint. 
Blended systems (those in category 4) are listed in the footnotes to
these tables. Atomic data were taken from \citet{Mo03}. 

\section{Sample statistics}
Among the DLAs that are unblended at \nf\ (those in categories 1, 2,
and 3), the overall \nf\ detection rate is 8/91 (secure detections
only), or 12/91 (secure and marginal detections). 
Applying small-number Poisson statistics \citep{Ge86},
and quoting 1$\sigma$ errors, the detection rate is therefore
9$^{+4}_{-3}$\% (secure) or 13$^{+5}_{-4}$\% (secure and marginal).
The \nf\ detection rate in sub-DLAs is 2/18 (11$^{+15}_{-7}$\%). 
In DLAs with unblended UVES data, the \nf\ detection rate is 
8/65 (12$^{+6}_{-4}$\%), 
and in DLAs with unblended HIRES data, the \nf\ detection rate is 
3/22 (14$^{+13}_{-7}$\%). Though this difference is not statistically 
significant, a marginally higher detection rate for HIRES is 
expected, because the median sensitivity of the HIRES spectra is slightly
higher: the median S/N per resolution element at \nf\ is 28.0, 39.8, and 37.7 
for the unblended data taken with UVES, HIRES, and ESI, respectively. 
Given the differences in pixel size, and assuming a line width
$b$=15~\kms, these S/N values correspond to limiting 
3$\sigma$ observer-frame equivalent widths at 4000\,\AA\ of 15.6~m\AA\ (UVES),
10.2~m\AA\ (HIRES), and 25.3~m\AA\ (ESI). 

As expected, the likelihood that the \nf\ data are blended is much
higher when the \nf\ doublet lies in the \lya\ forest. 
For a DLA at redshift \zabs, \nf\ $\lambda$1238 lies in
the \lya\ forest if 1238.8$\times$(1+\zabs)$<$1215.7$\times$(1+\zqso). 
Equivalently, \nf\ lies out of the forest if the DLA is within
5\,600~\kms\ of the background quasar, so for \nf, ``out of the forest''
also happens to imply ``proximate to the quasar''. 
Some 40$\pm$6\% (50/125) of all (intervening) 
DLAs in the \lya\ forest are blended at \nf, whereas none of the 16 
(proximate) DLAs lying out of the forest are blended at \nf.   
Interestingly, we find that once the blended systems are excluded,
the \nf\ detection rate for (intervening) DLAs in the forest 
is 10/75 (13$^{+6}_{-4}$\%) versus 2/16 (13$^{+16}_{-8}$\%) for
(proximate) DLAs 
outside the forest. The similarity of these detection rates suggests that
(a) blending has been correctly identified and no substantial biases
are introduced by rejecting cases where the \nf\ profiles are blended, and 
(b) there is no significant change in the likelihood of a \nf\
detection when the DLA is within $\approx$5\,000~\kms\ of 
the QSO, the traditional definition of ``associated'' or ``proximate'' systems.
We will return to this subject in \S6.

Note that the DLA \nf\ detection rate of 13$^{+5}_{-4}$\% contrasts with the 
100\% detection rate for \sif\ and \cf\ absorption in DLAs
\citep[][Paper II]{WP00a}, and the $>$34\% detection rate for \os\
absorption (Paper I; only a lower limit can be measured for \os, due to the
impact of blending with \lya\ forest lines). It also contrasts with
the very high detection rate of \nf\ in the Milky Way halo:
zero-redshift interstellar \nf\ absorption, though weak, is reported
in 32 of 34 Galactic halo sight-lines studied by \citet{IS04}.

\section{Neutral-phase elemental abundances}
\subsection{$\alpha$-elements and Zinc}
For each absorber in our sample, a [Z/H] and $N$(\hi) measurement is
available in the literature (see footnotes to Table A.1). 
For all systems with \nf\ detections, a measurement from
\ion{Si}{ii}/\hi\ is available. For the systems with \nf\
non-detections, we use either \ion{S}{ii}/\hi, \ion{Si}{ii}/\hi, or
\ion{Zn}{ii}/\hi, depending on the availability of unsaturated lines. 
S and Si (both $\alpha$-capture elements) and Zn (an iron-peak
element) are each undepleted onto dust grains, with a first
ionization potential $<$13.6\,eV, so that the singly ionized stage
is dominant in \hi\ regions. Studies have found that [Zn/$\alpha$]
in DLAs is rarely far from zero \citep{Ce00, Mo00, Ni04}, with
the latter authors finding [$\alpha$/Zn] $<$0.25 in ten DLAs covering
2\,dex of [Zn/H]. Thus within an error of $\approx$0.2 dex, the Zn abundance
can be used as a proxy for the $\alpha$-element abundance.
A potentially more serious problem is whether the metallicity
measurements are affected by ionization corrections.
For DLAs, photoionization simulations find that
ionization corrections are generally negligible \citep[][but see
  Howk \& Sembach 1999 and Izotov et al. 2001]{Vi95, Vl01}.
For sub-DLAs, \citet{DZ03} and \citet{Px07} report that ionization
corrections are typically small but depend on the element; 
\citet{Vl01} find that the relative ionization corrections at
log\,$N$(\hi)=20.2 are 0.15\,dex for O/Si, 0.23\,dex for O/S, and
0.64\,dex for O/Zn. The metallicities presented in Tables A.1 and A.2
have not been corrected for ionization.

\subsection{Nitrogen}
\begin{table*}
\begin{minipage}[t]{18cm}
\caption{Voigt component fits to DLA/sub-DLA \none}
\centering
\begin{tabular}{lc ccccc ccc}
\hline \hline
QSO & $z_{\rm abs}$\footnote{The DLAs at $z$=2.81115 toward \object{Q0528--2505}, $z$=1.96221 toward \object{Q0551--3637}, and $z$=3.09240 toward \object{J2100--0641} exhibit fully saturated \none\ profiles, and N abundances could not be extracted.} & log\,$N$(\hi) & Triplets\footnote{The triplets used to measure the \none\ absorption: \none\ $\lambda\lambda$ 1134.165, 1134.415, 1134.980 and \none\ $\lambda\lambda$ 1199.550, 1200.223, 1200.710.}  & $v_0$ & $b$ & log\,$N$ & [N/H]\footnote{Nitrogen abundance relative to solar where log\,(N/H)$_\odot$=$-$4.07 \citep{Ho01}. Errors were formed by adding in quadrature the logarithmic errors on $N$(\none), $N$(\hi), and an additional 0.05 dex allowing for uncertainties in the solar nitrogen abundance.} & [N/Si]\footnote{Nitrogen-to-silicon ratio.} & Ref.\footnote{All measurements in this table have been made by us. However, in several cases, [N/H] has been determined previously: 1 \citet{Le08}; 2 \citet{HP07}; 3 \citet{DZ06}; 4 \citet{DZ04}; 5 \citet{Lo02}; 6 \citet{Pe95}; 7 \citet{SP01}. Our results generally agree with previous results within the errors; for the DLA toward \object{Q2348-1444}, our value for log\,$N$(\none) is consistent with the upper limit log\,$N$(\none)$<$13.48 reported by \citet{Pe95}.}\\
    &           &   & & (\kms) & (\kms) & ($N$ in \sqcm) & & &\\
\hline
     J1014+4300 &                   2.95880 &  20.50$\pm$0.02 &    \none\ 1134, 1200 &     2$\pm$2 &  9$\pm$1 & 14.33$\pm$0.05 &                     $-$2.02$\pm$0.17 & $-$1.31$\pm$0.22 &   - \\
                &                           &                 &                      &    20$\pm$3 &  6$\pm$5 & 13.63$\pm$0.24 &                                                         &     \\
     J1211+0422 &                   2.37655 &  20.80$\pm$0.10 &    \none\ 1134, 1200 & $-$10$\pm$2 &  7$\pm$1 & 13.99$\pm$0.04 &                     $-$2.29$\pm$0.11 & $-$0.88$\pm$0.14 &   1 \\
                &                           &                 &                      &     0$\pm$2 &  4$\pm$1 & 13.85$\pm$0.07 &                                                         &     \\
                &                           &                 &                      &    14$\pm$2 &  7$\pm$1 & 14.04$\pm$0.03 &                                                         &     \\
     J1757+7539 &                   2.62530 &  20.76$\pm$0.02 &    \none\ 1134, 1200 & $-$21$\pm$3 & 17$\pm$2 & 14.57$\pm$0.08 &                     $-$1.61$\pm$0.07 & $-$0.82$\pm$0.07 &   2 \\
                &                           &                 &                      & $-$ 2$\pm$2 & 10$\pm$1 & 14.63$\pm$0.07 &                                                         &     \\
                &                           &                 &                      &    23$\pm$2 &  8$\pm$1 & 14.60$\pm$0.02 &                                                         &     \\
    Q0042--2930 &                   1.80947 &  20.40$\pm$0.10 &          \none\ 1200 &     9$\pm$2 &  8$\pm$2 & 14.06$\pm$0.05 &                     $-$2.26$\pm$0.12 & $-$1.01$\pm$0.19 &   - \\
    Q0450--1310 &                   2.06655 &  20.50$\pm$0.07 &    \none\ 1134, 1200 & $-$ 5$\pm$2 & 10$\pm$3 & 13.27$\pm$0.08 &                     $-$2.83$\pm$0.10 & $-$1.38$\pm$0.11 & 2,3 \\
                &                           &                 &                      &    26$\pm$3 & 14$\pm$3 & 13.32$\pm$0.07 &                                                         &     \\
     Q1242+0006 &                   1.82452 &  20.45$\pm$0.10 &    \none\ 1134, 1200 &     0$\pm$2 &  8$\pm$2 & 14.19$\pm$0.06 &                     $-$2.19$\pm$0.13 & $-$1.15$\pm$0.14 &   - \\
     Q1331+1704 &                   1.77635 &  21.15$\pm$0.07 &    \none\ 1134, 1200 &     6$\pm$2 & 13$\pm$1 & 15.06$\pm$0.06 &                     $-$1.89$\pm$0.10 & $-$0.49$\pm$0.11 & 2,4 \\
                &                           &                 &                      &    19$\pm$3 &  5$\pm$2 & 14.24$\pm$0.25 &                                                         &     \\
                &                           &                 &                      &    44$\pm$2 & 11$\pm$1 & 14.26$\pm$0.02 &                                                         &     \\
                &                           &                 &                      &    69$\pm$2 &  8$\pm$1 & 13.47$\pm$0.04 &                                                         &     \\
    Q2243--6031 &                   2.33061 &  20.65$\pm$0.05 &    \none\ 1134, 1200 & $-$ 1$\pm$2 &  7$\pm$1 & 14.59$\pm$0.03 &                     $-$1.99$\pm$0.08 & $-$1.14$\pm$0.09 & 2,5 \\
    Q2348--1444 &                   2.27940 &  20.56$\pm$0.08 &          \none\ 1200 & $-$ 2$\pm$2 &  4$\pm$2 & 13.06$\pm$0.09 &                     $-$3.43$\pm$0.26 & $-$1.51$\pm$0.26 & 2,6 \\
\hline
    Q0237--2321 &                   1.67234 &  19.65$\pm$0.10 &          \none\ 1200 &     1$\pm$2 &  6$\pm$2 & 13.39$\pm$0.06 &                     $-$2.19$\pm$0.25 & $-$1.93$\pm$0.27 &   - \\
    Q1037--2704 &                   2.13906 &  19.70$\pm$0.05 &    \none\ 1134, 1200 & $-$ 8$\pm$3 &  7$\pm$2 & 13.02$\pm$0.16 &                     $-$1.60$\pm$0.08 & $-$1.58$\pm$0.13 &   7 \\
                &                           &                 &                      &     7$\pm$2 &  9$\pm$2 & 13.55$\pm$0.05 &                                                         &     \\
                &                           &                 &                      &    43$\pm$2 &  8$\pm$1 & 13.78$\pm$0.02 &                                                         &     \\
\hline
\end{tabular}
\end{minipage}
\end{table*}
\normalsize

\none\ column densities were measured using the
\none\ $\lambda\lambda$1199.550, 1200.223, 1200.710 and
\none\ $\lambda\lambda$1134.980, 1134.415, 1134.165 triplets.
Simultaneous profile fits to whichever of these six \none\ lines were
unblended were executed to determine the best-fit \none\ column density.
The \none\ absorption-line spectra together with the best-fit models
are shown in Figures A.5 and A.6 in the Appendix. 
In three of the twelve DLAs with \nf\ 
detections (the DLAs at $z$=1.9622 toward \object{Q0551--3637}, $z$=2.8112
toward \object{Q0528--2505}, and $z$=3.0924 toward \object{J2100--0641}),  
we were unable to determine $N$(\none), due to complete saturation.
The latter two of these three are proximate DLAs.

In absorbers with log\,$N$(\hi)$\gg$19.5 (i.e. DLAs and strong
sub-DLAs), the \none/\hi\ ratio can be converted to 
nitrogen abundances without needing ionization corrections,
because charge-exchange reactions are fast enough to tie together the
ionization state of nitrogen and hydrogen \citep[and oxygen;][]{FS71, Vi95}. 
Thus we calculate the nitrogen abundance relative to solar as 
[N/H]=[log\,$N$(\none)--log\,$N$(\hi)]--log\,(N/H)$_\odot$, 
where log\,(N/H)$_\odot$=$-$4.07 is taken from \citet{Ho01}. 
The results are reported in Table 2. The assumption that
[\none/\hi]=[N/H] may break down if the ionization parameter $U\equiv
n_\gamma/n_{\rm H}$ is large enough \citep{Pr02b}, or for cases with
log\,$N$(\hi)$\approx$19.5. Thus for the two sub-DLAs in our sample
with log\,$N$(\hi)$\approx$19.7, [\none/\hi] may 
underestimate [N/H] because some of the N may exist in the
form of \ion{N}{ii}.

\section{Results}
Our \ndet\ \nf\ detections include two strong sub-DLAs
[each having log\,$N$(\hi)$\approx$19.70] and twelve genuine DLAs. 
They cover a redshift range from 1.67 to 3.09, and span a range of 
1.5~dex in neutral-phase metallicity [Si/H], from $-$1.92 to $-$0.44. 
Two of the DLAs with \nf\ are proximate, with one (the absorber at
$z_{\rm abs}$=3.0924 toward J2100--0641) 
lying at $\approx$3\,500\kms\ from the QSO redshift, and one 
(the absorber at $z_{\rm abs}$=2.81115 toward \object{Q0528--2505})
at $\approx$3\,200~\kms\ \emph{beyond} the QSO redshift \citep{SP98}.
The values of log\,$N$(\nf) among the detections lie between 13.20 and 
14.61, and the number of \nf\ components seen within a single DLA ranges
from one to eight. 

\subsection{Dependence of \nf\ detection rate on DLA properties}

\begin{figure*}[!ht]
\includegraphics[width=18.5cm]{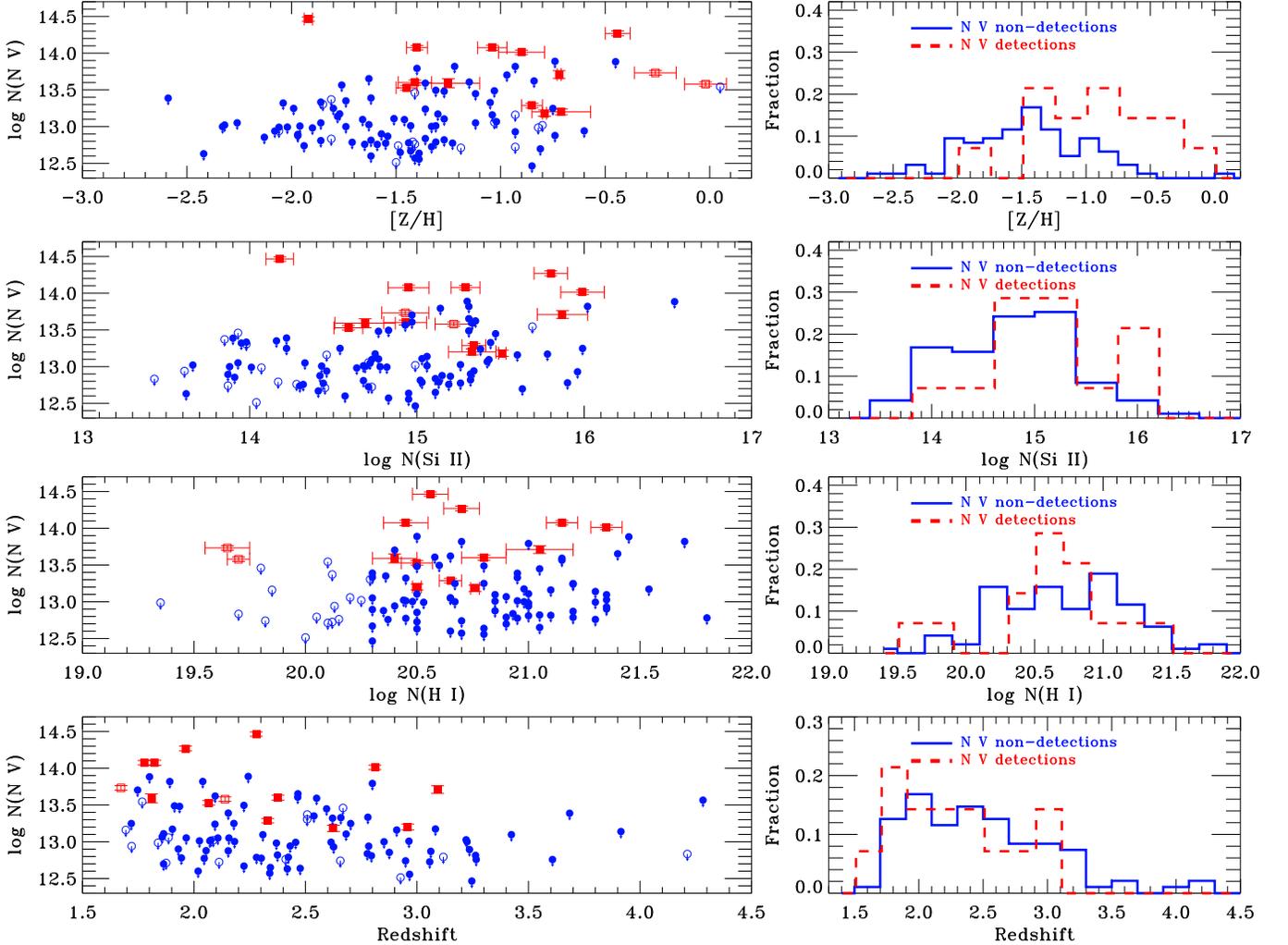}
\caption{Comparison of \nf\ column density with DLA neutral-phase
  metallicity (top), \ion{Si}{ii} column density (second
  panel), \hi\ column density (third panel), and absorber redshift
  (bottom) for DLAs (filled symbols) and sub-DLAs (open
  symbols). Detections are shown by red squares and non-detections by
  blue circles with upper-limit arrows.
  In each panel on the right, we show the distribution of the quantity
  plotted on the x-axis on the left, plotted separately for the \nf\
  detections and non-detections. Both sub-DLAs and DLAs are included
  in these distributions. The metallicity [Z/H] is 
  either \ion{Zn}{ii}/\hi, \ion{S}{ii}/\hi, or \ion{Si}{ii}/\hi.}
 \end{figure*}

The \nf\ column densities are plotted against known properties of the
DLA/sub-DLA ([Z/H], $N$(\ion{Si}{ii}), $N$(\hi), and $z$) in Figure 2,
to explore the factors that influence the formation of \nf.
We find that the \nf\ detections have neutral-phase
metallicities that distribute
very differently than the non-detections (top panel).
The median [Z/H] for the DLAs with \nf\ detections is $-$0.90, 
versus $-$1.46 for the DLAs with \nf\ non-detections. 
Three of the five highest metallicity DLAs in our sample, and
two of the three highest metallicity sub-DLAs, show \nf\ detections. 
A two-sided Kolmogorov-Smirnoff (K-S) test rules out the 
null hypothesis (the idea that the metallicity distributions of the
\nf\ detections and non-detections derive from the same parent
population) at the 99.2\% level. 
If the sub-DLAs are removed and the analysis is restricted to genuine
DLAs, the null hypothesis is again rejected, but with a slightly lower
confidence of 96.7\%. 
However, among the \nf\ detections, $N$(\nf) does
\emph{not} correlate with [Z/H].

We also find some evidence that the \ion{Si}{ii} column density
log\,$N$(\ion{Si}{ii})=log\,$N$(\hi)+[Z/H]+log(Si/H)$_\odot$
distributes differently for the \nf\ detections and non-detections
(second panel), but this difference is only significant at the
87.4\% level ($<$2$\sigma$). 
In other words, the probability of detecting \nf\ is somewhat 
sensitive to the total column of metals present in the neutral phase,
but is more sensitive to metallicity\footnote{Analogously,
  metallicity is a significant factor in determining whether
  H$_2$ is present in DLAs \citep{Pj06, No08}.}. 
The probability of \nf\ detection is not dependent on \hi\ column
density (third panel), and not strongly dependent on redshift (fourth panel).
However, we note there are no \nf\ detections at $z\!>\!3.1$, even
though the sample extends to $z$=4.28.

\begin{figure}[!ht]
\resizebox{\hsize}{!}{\includegraphics{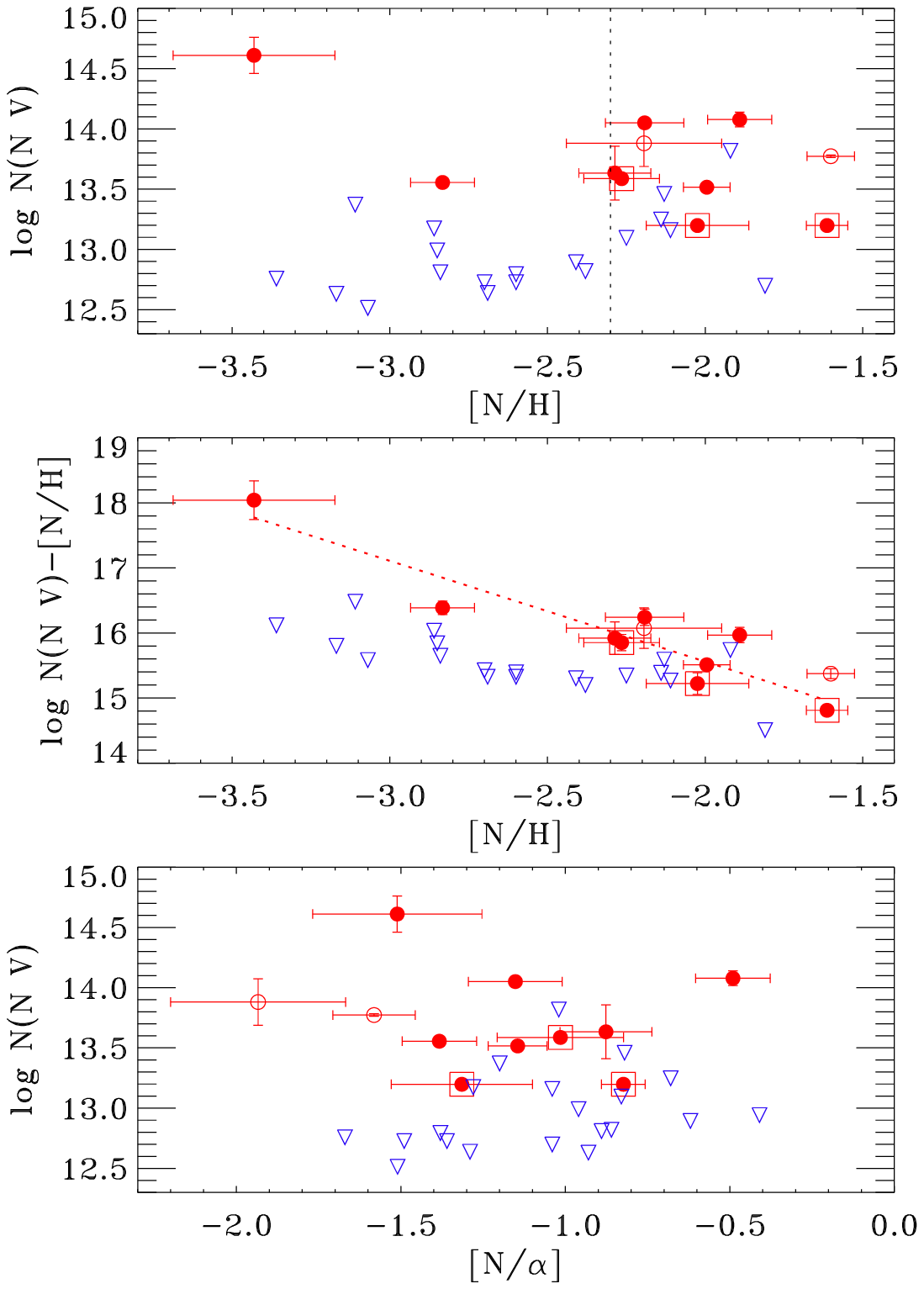}}
\caption{Dependence of \nf\ column density on neutral-phase nitrogen
  abundance and nitrogen-to-$\alpha$-element ratio. 
  The \nf\ detections (red) comprise nine DLAs (filled circles) and two
  sub-DLAs (open circles). The \nf\ non-detections (blue) have upper
  limits on $N$(\nf) from this paper and measurements of [N/H] and
  [N/$\alpha$] from the literature \citep{Pj08, Pe08}. Marginal
  \nf\ detections are highlighted with open squares. The
  $\alpha$-element is either Si or S. 
  The dashed line in the top panel
  shows [N/H]=$-$2.3, where a change in the \nf\ detection probability
  is observed. The solid line in the middle panel shows a linear fit
  to $N$(\nf)--[N/H] vs [N/H] for the \nf\ detections only.} 
\end{figure}
 
In the sub-sample of DLAs/sub-DLAs with \nf\
detections/upper limits \emph{and} \none\ measurements, we investigate in
Figure 3 whether the \nf\ column density correlates with 
neutral-phase nitrogen abundance and nitrogen-to-alpha ratio. We
find that \nf\ is much more likely to be detected in the systems with [N/H]
above $-$2.3 (top panel): applying Poisson statistics, \nf\ is
detected in 57$^{+22}_{-16}$\% of systems at [N/H]$>\!-$2.3 but only
13$^{+18}_{-9}$\% of systems at [N/H]$<\!-$2.3. 
Of course, one expects that the \nf\ detection rate will increase with
[N/H], because for a fixed total plasma column and nitrogen ionization
fraction, $N$(\nf) is proportional to [N/H]. However, 
among the \nf\ detections, log\,$N$(\nf) does not correlate with
[N/H]. Equivalently, if we correct for the higher 
detectability of the higher [N/H] systems by plotting $N$(\nf)/[N/H] versus
[N/H] (middle panel) for the detections, we find that these two
quantities are anti-correlated. This surprising result
implies that either (a) the
higher-metallicity systems show \emph{smaller} total plasma columns 
than the lower-metallicity systems (the plasma column $N$(\hw) is
$\propto$ $N$(\nf)/(N/H) for a fixed ionization fraction), or (b) the
assumption that the neutral and plasma phases share a common
nitrogen abundance is false, or (c) the assumption that
[N/H]=[\none/\hi] in \nf-bearing DLAs is false.
Given the presence of three marginal \nf\ detections, two sub-DLAs (whose
metallicity measurements may be effected by ionization corrections),
and one absorber with very unusual properties (the DLA toward
\object{Q2348-1444}), it would be premature to over-interpret this
trend by favoring one of these three alternatives. 
Nonetheless, the observation that the DLAs with larger
[N/H] do not show proportionally larger $N$(\nf) implies that we can rule
out a model in which neutral-phase nitrogen abundance is the only
factor influencing the strength of \nf. 
Finally, no correlation is observed between
$N$(\nf) (or \nf\ detection rate) and the neutral-phase [N/$\alpha$]
ratio (lower panel): we find a dispersion of over 1\,dex in $N$(\nf)
at a given [N/$\alpha$].

\begin{figure}[!ht]
\resizebox{\hsize}{!}{\includegraphics{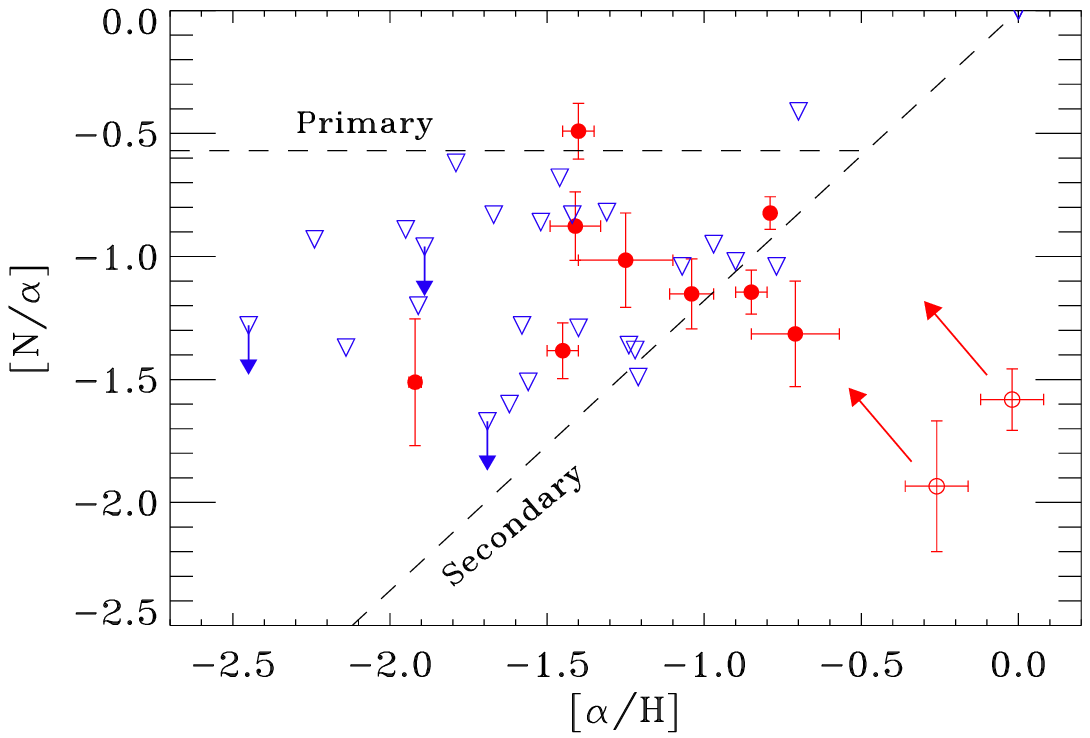}}
\caption{Comparison of neutral-phase abundance patterns
  for systems with and without \nf\ detections.
  The \nf\ detections (red) include nine DLAs (filled circles) and two
  sub-DLA (open circles). 
  The \nf\ non-detections (blue open triangles) have limits on
  $N$(\nf) from this paper and [$\alpha$/H] 
  taken from either \citet{Pe08} or \citet{Pj08}.  
  The $\alpha$-element is either Si or S.
  The effect of ionization
  corrections on the two sub-DLAs is shown with arrows.
  The dashed lines show the local primary plateau at [N/O]=$-$0.57
  \citep{Na06}, and the locus of secondary nitrogen abundance, formed by
  extrapolating local measurements to low O/H \citep[see][]{Pj08}.}
\end{figure}

Six of the nine DLAs with \nf\ and \none\ measurements fall in the 
region between the primary and secondary plateaus\footnote{Cosmic nitrogen
production is thought to be dominated by the CNO cycle in
intermediate-mass stars; primary nitrogen is synthesized from
``seed'' carbon and oxygen atoms that themselves were produced in
the same stars; secondary nitrogen is synthesized from carbon
and oxygen atoms that were present when the star formed.} in a plot
of [N/$\alpha$] vs [$\alpha$/H] (Figure 4).
Two of the remaining three lie slightly below the secondary plateau
\citep[as defined by][]{Pj08}, with low N/$\alpha$ for their
$\alpha$/H, and the third shows [N/$\alpha$] marginally above the
primary plateau, but overall, there is nothing particularly unusual
about the location in this plane of the DLAs with \nf\ detections.
The two sub-DLAs with \nf\ fall in a different region on Figure 4
than the DLAs, each showing extremely low [N/$\alpha$] for their
[$\alpha$/H]\footnote{In the sense of showing low N/$\alpha$ for their
  $\alpha$/H, the two sub-DLAs in our sample behave similarly to cB58,
  a case-study for Lyman Break Galaxies (LBGs) at $z$=2.7 \citep{Pe02b}.}. 
Although it is tempting to interpret this as evidence that sub-DLAs
represent a fundamentally different category of absorber than DLAs
\citep[indeed, many studies have found sub-DLAs have higher mean metallicities;
][]{Px07, Px08, Ku07, Me07, Me08}, we cannot rule out the possibility
that in these two cases, we are seeing ionization effects. Indeed,
this would cause the sub-DLA [N/H] and [N/$\alpha$] to be
\emph{underestimated} and [$\alpha$/H] to be \emph{overestimated}. 
Correcting for these effects would bring the sub-DLAs points back
toward the DLA points. 
Furthermore, our selection by \nf\ (i.e. by the presence of
highly-ionized plasma) implies that these DLAs and sub-DLAs are the
very absorbers where ionization effects are important to begin with. 

\subsection{Relationship between \nf\ and other high ions}
In this sub-section we consider the relationship between the
\nf-absorbing plasma in DLAs and the phase(s) traced by the other high ions
(\sif, \cf, and \os). First, we compare to \os.  
Of the twelve DLAs in our sample with \nf, four show \os\ detections 
(the DLAs toward \object{Q0450--1310}, \object{Q0528--2505},
\object{Q2243--6031}, and \object{J1211+0422}); in the other eight
cases the \os\ profiles are blended or not covered.
Among the two sub-DLAs with \nf\ detections, one (toward
\object{J0240--2309}) has no coverage of \os, and the other (toward
\object{Q1037--2704}) shows strong (and partially blended) \os\ absorption.
More straightforward is a comparison with \cf, which (like \sif) is
detected in all the DLAs and sub-DLAs with \nf.
A clear correlation exists between the total \cf\ and
\nf\ column densities (integrated over components), as shown in Figure
5, top panel. All except one of the \nf\ detections have
log\,$N$(\cf)$\ga$14.6 (and the exception is a marginal \nf\ detection). 
The median log\,$N$(\cf) in DLAs and sub-DLAs with secure \nf\
detections is 15.16, whereas the median log\,$N$(\cf) in 
the general DLA population is 14.2 (Paper II). 
In the lower panel of Figure 5, we find a good correlation between
$N$(\nf) and the total velocity width of \cf\ absorption $\Delta v$ 
(defined as the velocity width containing the central 90\% of the
integrated optical depth). All-bar-one of the DLAs and sub-DLAs with
\nf\ show \cf\ absorption extending over at least 200~\kms, 
and half show \cf\ extending over at least 350~\kms.
Thus \nf\ is preferentially formed in the absorbers with the most
extended high-ion kinematics. We note that while these correlations show that
the \cf\ and \nf\ are closely related, they do not necessarily
indicate that the two ions are co-spatial.

\begin{figure}[!ht]
\resizebox{\hsize}{!}{\includegraphics{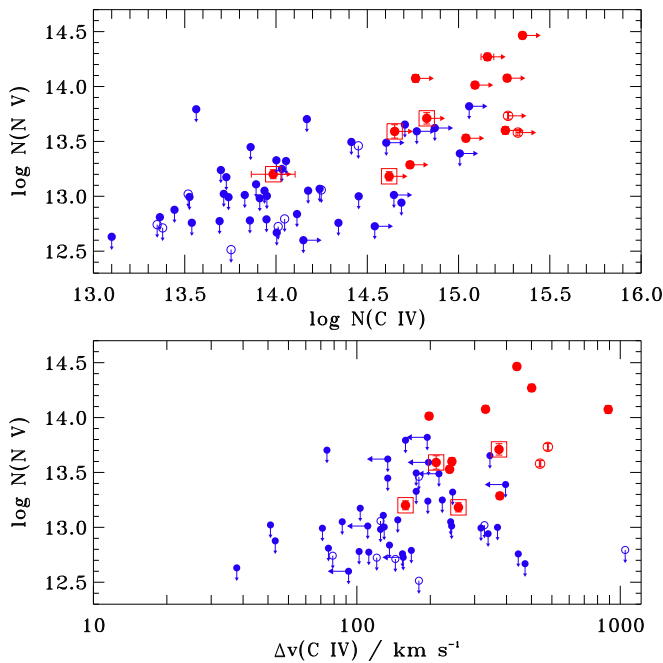}}
\caption{Comparison of \nf\ column density with 
  \cf\ column density (top) and total \cf\ line width (bottom), for
  DLAs (solid circles) and sub-DLAs (open circles). 
  \nf\ detections are shown in red, and \nf\ non-detections are
  shown in blue with downward-pointing upper-limit arrows.
  The marginal \nf\ detections are highlighted with open squares. 
  Horizontal arrows indicate DLAs with saturated \cf\ lines. 
  \cf\ measurements are taken from Paper II.}
\end{figure}

To address directly whether two given ions are co-spatial, one 
needs to compare the optical depth profiles of unsaturated transitions. 
Because of the relatively low $f$-values of the \nf\ doublet
combined with the low nitrogen abundance in DLAs, this comparison is
not always possible, since the other high-ion lines tend to be
saturated at the velocities where \nf\ is detected. However, in eight
DLAs and two sub-DLAs, the \nf\ absorption is accompanied by
unsaturated absorption in at least one of \cf\ $\lambda$1548, \cf\
$\lambda$1550, \sif\ $\lambda$1393, and \sif\ $\lambda$1402, so that a
apparent optical depth \citep[AOD;][]{SS91} comparison can be made.
This comparison is shown in Figure 6. We find that in
four out of five cases where we can compare the \nf\ and \cf\
profiles, they coincide closely, which (given the similar atomic
weights of nitrogen and carbon) suggests co-spatiality of the two ions.
The exception is the marginal \nf\ detection in the
DLA toward \object{Q0042--2930}, where the \nf\ profile is displaced
by $\approx$4~\kms\ relative to the \cf.
However, in four out of five cases where we compare \nf\ and \sif, clear
\emph{differences} between the profiles are observed: the \sif\ 
profiles tend to be more complex, and contain narrower components,
than the \nf\ profiles.

\begin{figure*}[!ht]
\includegraphics[width=18.5cm]{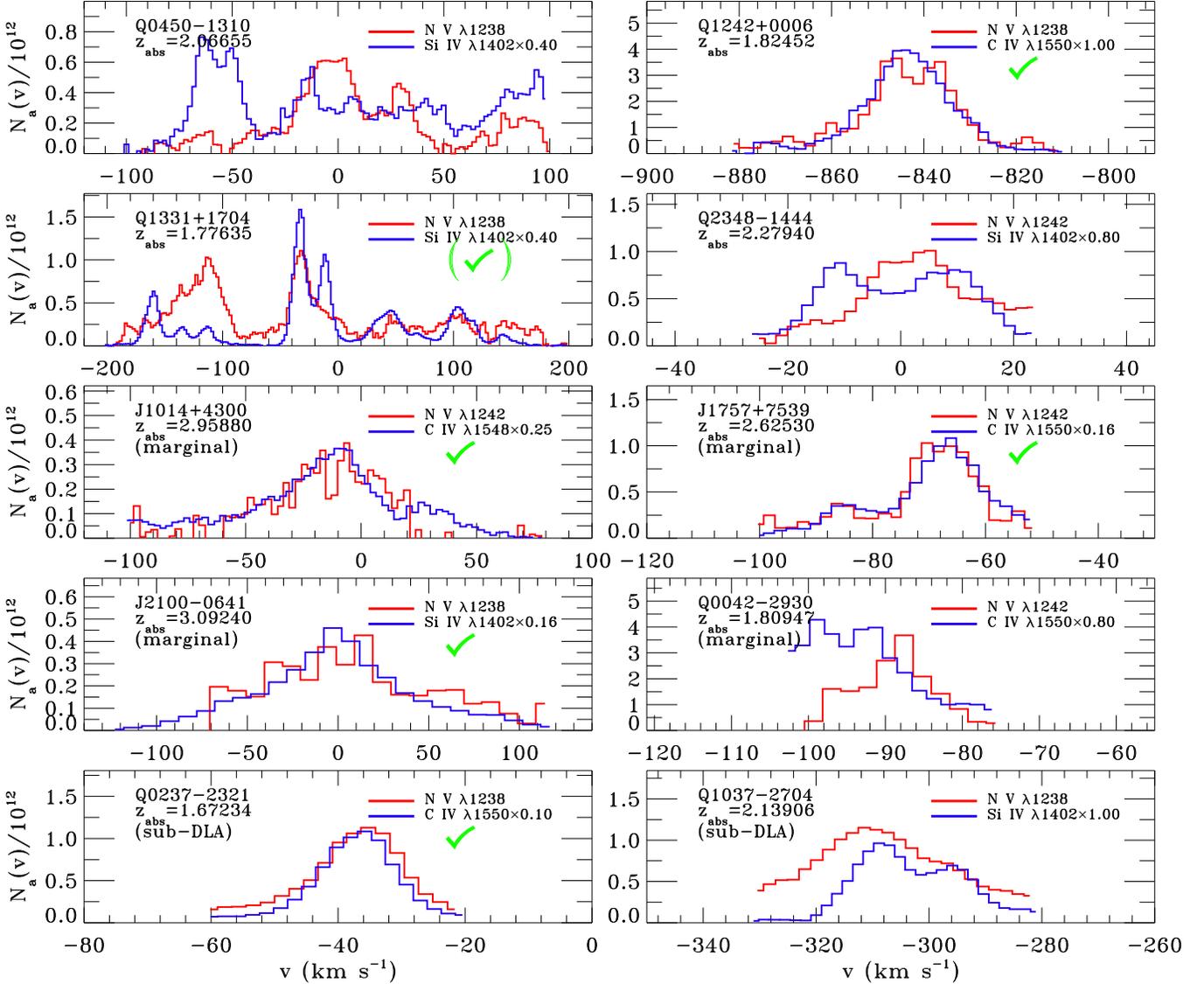}
\caption{Comparison of apparent column density profiles of \nf\ with
  either \cf\ or \sif, in cases where the absorption in the comparison
  line is unsaturated. The unit on the y-axis is
  ions~cm$^{-2}$~(km\,s$^{-1}$)$^{-1}$. The \cf\ and \sif\ profiles
  have been scaled by the factors annotated on the panels for ease of
  comparison. A green tick mark indicates that the \nf\ and the
  comparison ion show similar profiles (overlapping within the errors). 
  A tick mark in parentheses indicates the profiles coincide over part
  of the velocity range, but not in other regions.} 
\end{figure*}

We can further investigate whether the
high ions trace similar temperature regions by analyzing the component
line width distributions. The component line width (Doppler
$b$-parameter) provides information
on the temperature of the absorbing gas, according to
$b=\sqrt(\frac{2kT}{Am_{\rm H}}+b_{\rm non-th}^2)$ 
where $A$ is the atomic weight and $b_{\rm non-th}$ is the
non-thermal contribution to the line width. 
Since the atomic weights of oxygen (16), carbon (12), and nitrogen (14)
are similar, the thermal line widths for \os, \cf, and
\nf\ lines formed in the same region of gas are similar, 
though the \sif\ lines will be narrower, since $A$(silicon)=28. 
The presence of any non-thermal 
broadening will further reduce the differences between $b$(\os), $b$(\cf),
and $b$(\nf). Therefore, a $b$-value analysis can
test whether the various high ions are co-spatial.
Such an analysis is shown in Figure 7. In the top panel, close
similarities are seen between the distributions of $b$(\sif),
$b$(\cf), and $b$(\nf), which show the median values of 13.1, 16.3,
and 18.3~\kms, respectively. 
Although there is a slight excess of \nf\ components at $b$=18--24~\kms\
(relative to \cf), we cannot rule out the idea that the $b$-value
distributions for \nf\ and \cf\ are drawn from the same parent
population. This further supports the idea that these two ions 
trace the same regions of gas. Note that one can always add extra
components to any Voigt-profile fit, and the effect of this is to
shift the distribution of line widths to narrower values. However,
since we use the same fitting procedure for all high ions, this effect
should not change our conclusions.

\begin{figure}[!ht]
\resizebox{\hsize}{!}{\includegraphics{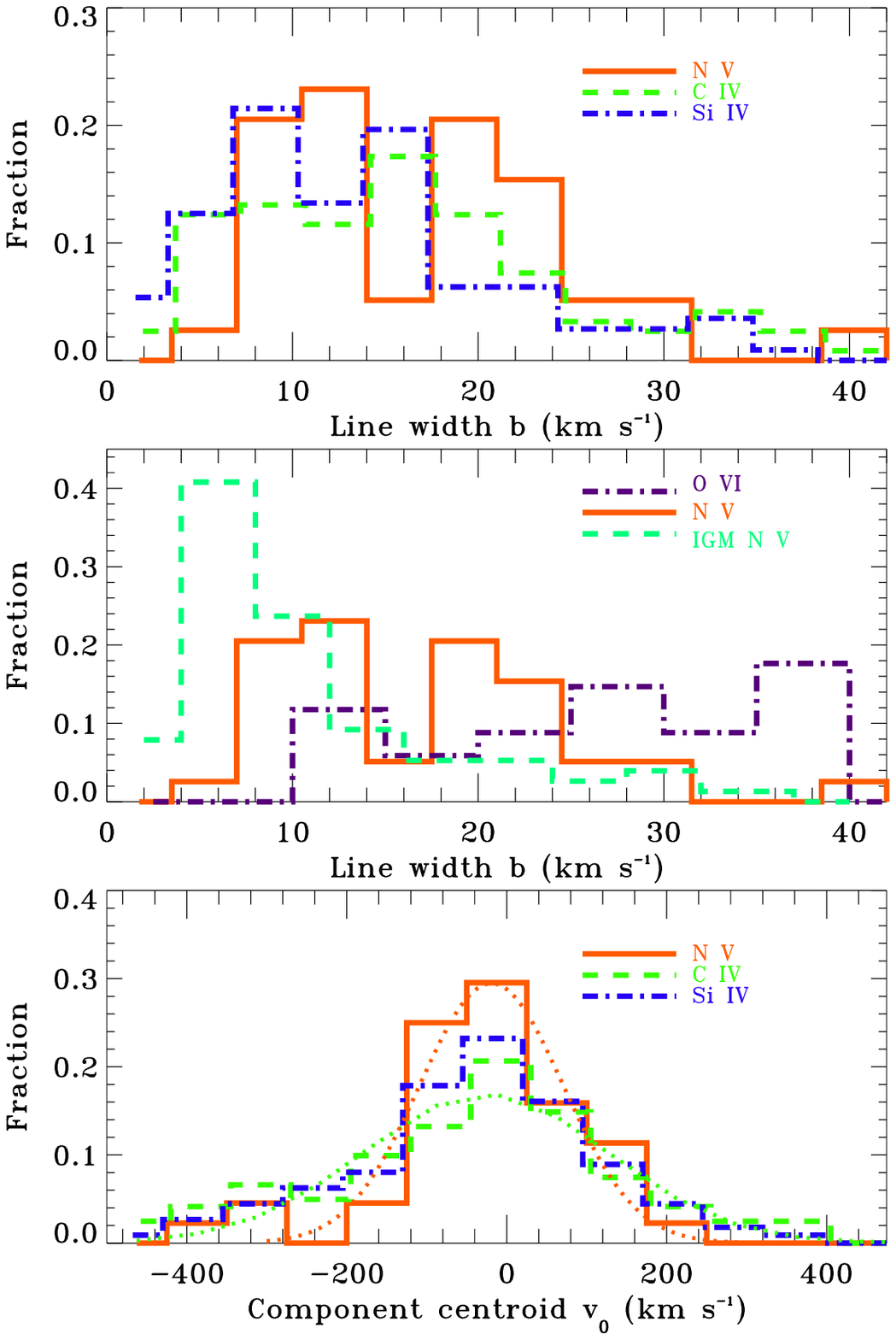}}
\caption{Histograms of $b$-value and velocity centroid $v_0$
  (defined relative to \zabs) found by VPFIT for the \ncomp\ \nf\ 
  components identified in the twelve DLAs and two sub-DLAs with \nf. 
  In the top and bottom panels, corresponding distributions for
  the 121 \cf\ and 112 \sif\ components \emph{in the 
  same absorbers} are shown. In the middle panel, we 
  include the $b$-value distribution of the \os\ components in a 
  separate sample (with some overlap) of 12 DLAs in Paper I and 
  the distribution of \nf\ $b$-values in the IGM sample of
  \citet{FR09}. The dotted lines in the bottom panel show Gaussian
  fits to the distribution of \cf\ and \nf\ component centroids.
  Four fitted \nf\ components with $b\!>\!45$~\kms\ and two with
  $|v_0|\!>400$~\kms\ are not included in the distributions.}
\end{figure}

Whereas the \nf\ component line width distribution closely follows the
\cf\ and \sif\ distributions, it differs from the \os\ distribution
(middle panel) taken from Paper I.
The \os\ distribution has a broader median $b$-value of 25.0~\kms, 
whereas 86\% of the \nf\ components have $b\!<\!25$~\kms. 
A two-sided K-S test finds 99.9\% evidence that that the \os\ and \nf\
components are not drawn from the same parent population. 
Thus we conclude that the majority of \nf\ components in DLAs arise in 
different (cooler) regions than the \os\ components.
However, a word of caution is necessary, because the \nf\ and \os\ 
lines are detected in largely different samples of absorbers
(only four DLAs show both \nf\ and \os). 
That is, we have not shown that the \os\ components are broader than
the \nf\ components \emph{within individual systems}; rather,
we have shown than the median \os\ line width in those DLAs where we
can detect and measure the \os\ is broader than the median \nf\ line
width in those (mostly different) DLAs where we can detect and measure
the \nf. 

Another result visible on Figure 7 (middle panel) is that the median
\nf\ $b$-value in DLAs and sub-DLAs of 18.3~\kms\ is significantly
broader than the median \nf\ $b$-value of 6.0~\kms\ measured by
\citet{FR09} in their survey for (intergalactic) \nf\ absorption at
high redshift. We discuss the implications of this in \S6.2. 
In the bottom panel of Figure 7, we note that the velocity centroids
($v_0$, defined relative to \zabs) distribute similarly for \cf\ and
\sif, but the \nf\ distribution is narrower.  A Gaussian fit to the
\cf\ component centroid distribution finds a FWHM of 361$\pm$38~\kms,
whereas for \nf, the FWHM is 207$\pm$19~\kms. 
This difference is due to sensitivity: the detectability of \nf\
depends on the total plasma column density, so low-column-density,
high-velocity satellite components will not be seen (particularly if
their metallicity is low).
 
\subsection{The \nf\ ionization fraction}
The \nf\ ionization fraction $f$(\nf)$\equiv$\nf/N is an important 
quantity for calculating the total ionized column density
in a given absorber (see \S5.4), since it quantifies the amount
of nitrogen present in unseen ionization stages. $f$(\nf) is
temperature-dependent in the case of collisional ionization, and
density-dependent in the case of photoionization. In this sub-section
we investigate the allowed range of values $f$(\nf) can take
in a wide range of parameter space.

\begin{figure}[!ht]
\resizebox{\hsize}{!}{\includegraphics{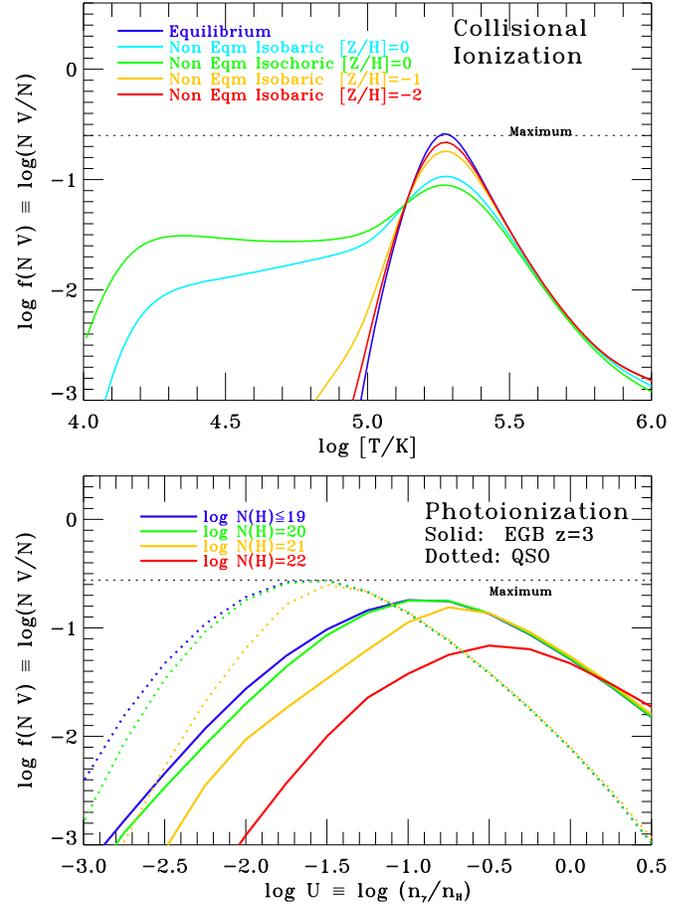}}
\caption{Top panel: dependence of the \nf\ ionization fraction
  $f$(\nf) on temperature in the
  collisional ionization models of \citet{GS07}. Both equilibrium- and
  four non-equilibrium models are shown, covering a range of
  metallicity from solar to one-hundredth solar. 
  Bottom panel: dependence of
  $f$(\nf) on ionization parameter $U\!\equiv\!n_\gamma/n_{\rm H}$
  in our set of CLOUDY photoionization models, which simulate gas
  illuminated by the extragalactic background at $z$=3 (solid lines)
  or with an unattenuated QSO power-law spectrum (dotted
  lines). Different colors are used to show clouds of different total
  hydrogen column density $N$(H)=$N$(\hi+\hw). }
\end{figure}

The predictions for $f$(\nf) for a range of equilibrium and
non-equilibrium \emph{collisional} ionization models from \citet{GS07} are
shown in the upper panel of Figure 8. We display models 
computed with solar elemental abundances, and also with one-tenth and
one-hundredth solar. For the solar abundance models, we show both the
isobaric and isochoric cases, but for the 0.1 and 0.01 solar models,
the isobaric- and isochoric-case predictions for $f$(\nf) are very
similar, so we only show the isobaric models. 
It can be seen that the maximum value for $f$(\nf) is 0.25 ($-$0.6 in
the log), in any of the collisional ionization models. Note also that
in the solar-metallicity non-equilibrium models, \nf\ can exist at
detectable levels down to surprisingly low temperatures
(log\,$T$=4.2--5.0). Thus a narrow \nf\ component does not
necessarily have to arise in photoionized plasma; it can also arise in
collisionally-ionized plasma that has cooled out of equilibrium. However, a
``cooled-gas'' scenario in which collisionally-ionized \nf\ is present
at temperatures of log\,$T$=4.2--5.0 is only possible for the solar
metallicity case; for lower metallicities, as would apply in the DLAs
under study here, the non-equilibrium models follow the equilibrium
models closely and predict essentially no \nf\ at log\,$T\!<5.0$.

To investigate the range of values that $f$(\nf) can take in
photoionized plasma, we conducted a series of CLOUDY (v08.00) models.
CLOUDY \citep[last described in][]{Fe98} is a plane-parallel radiative
transfer code which can predict the complete ionization breakdown in a
volume of gas of given density and abundance exposed to a given
radiation field. 
We took two cases for the input radiation field, one representing the
extragalactic background (EGB) at $z$=3 calculated by \citet{HM96}, 
and another representing an unattenuated power-law QSO spectra with
the form $F_\nu\propto\nu^{-1.5}$ at $\lambda\!<\!1216$~\AA.
For each radiation field, we ran a grid of models at different ionization
parameter log\,$U$, where $U$ is the ratio of ionizing photon density
to gas density, in each case computing the value of $f$(\nf). We then
repeated this procedure for clouds of different total hydrogen column
density $N$(H)=$N$(\hi+\hw).
We assumed [Z/H]=$-$1.5 for all CLOUDY runs, with a solar relative
abundance pattern maintained, though the choice of [Z/H] does not
affect $f$(\nf) (see below). 

The results for $f$(\nf) in the photoionization case are shown in the
lower panel of Figure 8.
We find that for log N(H)$\la$19, $f$(\nf) is essentially
independent of N(H); this is the optically thin case.
For the EGB runs, we find that $f$(\nf) is maximized for values of
log\,$U$ between $-$1.2 and $-$0.6, corresponding (using the EGB flux
density at $z$=3) to gas densities log\,[$n$(H)/cm$^{-3}$] between
$-$3.6 and $-$4.2, 
but $f$(\nf) never exceeds an absolute value of 0.20
($-$0.7 on the log scale).
For the runs with the QSO radiation field, $f$(\nf) is maximized at
lower values of log\,$U$, peaking between log\,$U$=$-$2.0 and $-$1.0,
but does not exceed 0.25 ($-$0.6 in the log).
The choice of [N/H] is unimportant: we found that the \nf\
ionization fraction is insensitive to the absolute or the relative
N abundance. When repeating the runs with the EGB at $z$=2
rather than $z$=3, we found essentially identical results. 

In summary, $f$(\nf)$<$0.25 in a range of collisional ionization and
photoionization models covering a large range of parameter space. This
justifies our use of a minimum correction factor of four when converting
the \nf\ column density to the total nitrogen column density.

\subsection{\hw\ column density}
A key goal in the study of DLA galaxies
is to determine the relative fractions of neutral and ionized gas. 
Following the approach described in Paper I, we can calculate
the \hw\ column density contained in the \nf\ absorbers by correcting
for ionization and metallicity:
\begin{equation}
N(\hw)=\frac{N(\nf)}{f(\nf)\rm{(N/H)_n}}
\frac{\rm{(N/H)_n}}{\rm{(N/H)_i}}, 
\end{equation}
where (N/H)$_{\rm n}$ and (N/H)$_{\rm i}$ denote the nitrogen
abundances in the neutral and ionized phases, respectively. 
Only (N/H)$_{\rm n}$ can be measured directly, 
and we assume equal abundances in the neutral and highly-ionized
phase, i.e. (N/H)$_{\rm n}$/(N/H)$_{\rm i}$=1.
We adopt $f$(\nf)$<$0.25 (see \S5.3), to give a \emph{lower} limit on
$N$(\hw). $N$(\hw) represents the H$^+$ column living with the \nf;
it does not account for H$^+$ existing at other temperatures.
$N$(\hw) is plotted against $N$(\hi) in Figure 9,
where we also show the implied \hw\ column densities in 
twelve DLAs (Paper I) and three sub-DLAs \citep{Fo07c} with \os\
detections.

\begin{figure}[!ht]
\resizebox{\hsize}{!}{\includegraphics{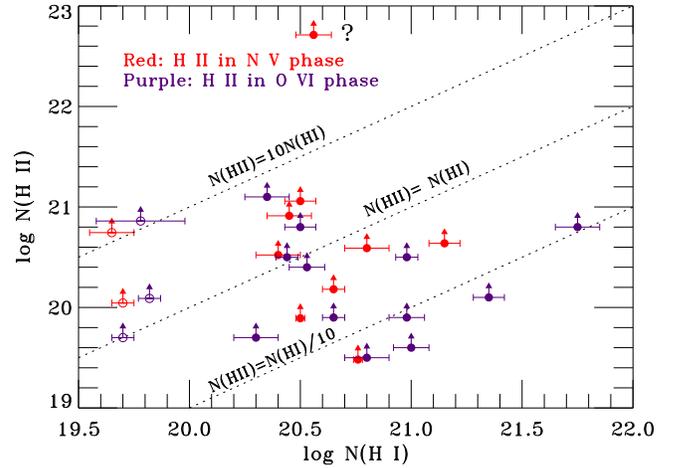}}
\caption{Comparison of hydrogen column densities in the highly-ionized  
  and neutral phases of DLAs (filled circles) and sub-DLAs (open
  circles), 
  for systems with \nf\ (red) and \os\ (purple) detections. 
  $N$(\hw) in the \nf\ (\os) phase is calculated assuming the same 
  N (O) abundance as in the neutral gas.
  All data points are lower limits since we have used the most
  conservative (smallest) ionization corrections allowed 
  [$f$(\nf)$<$0.25 and $f$(\os)$<$0.20].
  From top to bottom, the three dotted lines show the
  loci of DLAs with [10, 1, 1/10] times as much highly-ionized- as
  neutral gas. 
  The data point marked '?' is discussed in the text.}
 \end{figure}

We find that when \nf\ is detected in a DLA, the implied column of
ionized hydrogen [log\,$N$(\hw)$\sim$19.5--21.0]
is of similar order to the column contained in the \os\ phase of DLAs.
The median lower limit on $N$(\hw)/$N$(\hi) 
among the nine intervening DLAs with \nf\ and \none\ measurements is 0.34, 
whereas in Paper I we found a median lower limit on $N$(\hw)/$N$(\hi)
of 0.40 among nine intervening DLAs with \os.  
However, only 13$^{+5}_{-4}$\% of DLAs show \nf, whereas at least 34\%
show \os, so 
globally \os\ is a better tracer of the highly-ionized baryons and metals.
The lower limits on $N$(\hw) implied by \nf\ take similar values in
the sub-DLAs as in the DLAs. However the ratio \hw/\hi\ is much higher in the
sub-DLAs, since their $N$(\hi) values are lower. We find 
$N$(\hw)/$N$(\hi)$>$1 in all four sub-DLAs with \os\ or \nf, and $>$10
in two of these cases\footnote{Whether or not the \hw\ columns derived
  from \nf\ and \os\ should be summed depends on whether the two lines
  are co-spatial. Further line-profile comparisons are needed
  to investigate this.}.
Furthermore, in sub-DLAs the \hi\ phase itself is
thought to be ionized: \citet{Px07} report a mean ionization fraction
of 68\% in a sample of 26 sub-DLAs with photoionization
models. Therefore, a double ionization correction is necessary to
determine the total baryon content in a sub-DLA system \citep{Fo07c}.

One system bears mentioning in more detail: the DLA at \zabs=2.2794
toward \object{Q2348--1444}. The extremely strong high-ion absorption
lines observed in this DLA are more typical of quasar-intrinsic
``associated'' absorption-line systems rather than intervening absorbers. 
If the assumption of equal metallicity in the neutral and \nf-phases
is correct, this DLA has an extremely high implied $N$(\hw), over 100
times larger than the \hi\ column (this data point is marked ``?''
near the top of Figure 9). However, if abundance inhomogeneities caused the
true value of (N/H)$_{\rm n}$/(N/H)$_{\rm i}$ to be $<$1,
the value of $N$(\hw) would be reduced accordingly.
This could occur if, for example, the strong \nf\ lines were
formed in a nitrogen-enriched \hw\ region around an AGB star, while the
\none\ and \lya\ lines trace the general, metal-poor ISM. Another
possibility is that in this DLA, the \lya\ lines are not formed in
neutral gas, but rather in a series of closely separated,
predominantly ionized clouds whose total \hi\ column sums to over
2$\times$10$^{20}$~\sqcm. In this scenario, the measured extremely low
nitrogen abundance is an artifact caused by most of the N existing in
the form of \ion{N}{ii}, but the true nitrogen abundance is higher,
which in turn would lower the estimate of log $N$(\hw) in the \nf\
phase.

The large column densities of plasma contained in the \nf\ phase
of DLAs argue against photoionization by the EGB, for the following reason.
If we take the representative values log\,$U$=$-$1.0 and
log\,$n$(H)=$-$3.8 from the EGB model (\S5.3), and the typical total
\hw\ column density in the \nf\ phase of log\,$N$(\hw)=20.5 (Fig. 9),
we derive a path length (cloud depth along the line-of-sight) of 
$l=N/n=10^{24.3}$~cm or $\sim$650~kpc. Given the expected Hubble
parameter at $z$=2.5 of 250~\kms~Mpc$^{-1}$ \citep[from a standard
WMAP cosmology;][]{Sp07}, such a cloud would be broadened to a line width of
$\sim$160~\kms, {\it almost a factor of ten larger than the observed median
\nf\ component line width of 18~\kms}.
If log\,$n$(H) were higher or lower than the value of $-$3.8, the
\nf\ ionization fraction decreases and $N$(\hw) increases, making the
problem worse. Even taking a conservatively low value
log\,$N$(\hw)=20.0, we still arrive at a path length a factor of
$\approx$3 too large. Therefore, photoionization of \nf\ in DLAs
by the EGB can be ruled out \citep[this conclusion was also reached
  by][]{Le08}; however, we cannot rule out photoionization by local
sources.

\section{Discussion} 
High-redshift ($z\!\approx\!2$--3) \nf\ absorption is found in several
environments, including the intergalactic medium 
\citep{Be02, Ca02, Si02, Bo03, Si06, Sy07, FR09}, 
quasar-intrinsic or ``associated'' systems
\citep{Ha97, PS99, DO04, Fe04, FR09, Fo08a},  
LBGs \citep{Pe02b, Sh03}, and gamma-ray burst (GRB) host galaxies
\citep{Pr08c, Fo08b, Th08}. As for the LBG and GRB cases,
the DLA/sub-DLA \nf\ detections reported in this paper sample
\emph{galactic} (or circumgalactic) plasma. A key difference is that 
the DLA/sub-DLA sample is not selected by luminosity, but rather by
the presence of neutral gas. Consequently, the DLA/sub-DLA host
galaxies likely cover a range of halo masses and morphological types. 

\subsection{The \nf\ detection rate in DLAs}
We report a strong dependence of the \nf\ detection rate on
neutral-phase metallicity, and in particular on nitrogen abundance,
finding that the \nf\ detection probability quadruples above 
[N/H]=$-$2.3. However, among the detections, $N$(\nf) does not
correlate with [N/H] or with [$\alpha$/H]. In this sense, \nf\ differs from
\cf\ and \os, which each show strong correlations with [Z/H] (Paper I;
Paper II). This difference may relate to the unusual nucleosynthetic
history of nitrogen. Nonetheless, \nf\ 
is preferentially found in DLAs with the largest high-ion
velocity widths (Figure 5). Such extended high-ion absorption 
has been associated with feedback-driven galactic winds (Paper
II). This association is supported by several galactic wind simulations 
\citep{OD06, KR07, Fa07}, which all predict significant high-ion absorption.
Furthermore, a scenario where highly-ionized plasma in DLAs is
produced following star formation (providing the metals) and Type II
supernovae (providing the heating and the extended kinematics)
naturally explains the $N$(\cf)-[Z/H] and $N$(\os)-[Z/H] correlations.

We observe no DLAs with \nf\ at $z\!>\!3$, even though the sample
extends to $z$=4.28. This can be explained by the $z\!>\!3$ systems
tending to have lower metallicity, as well as the increasing density
of the \lya\ forest making high-$z$ \nf\
detections more challenging\footnote{A changing UV background following
  \ion{He}{ii} reionization, thought to occur at
  $2.9\!\la\!z\!\la\!3.4$ \citep{Re97, Hp00, Th02, Zh04}, could also
  explain this trend, but it would be premature to invoke this given
  the other explanations.}.
Interestingly, the detection rate of \nf\ in our proximate DLA sample 
(13$^{+18}_{-9}$\%) is the same as for our intervening DLA sample
(13$^{+5}_{-4}$\%). 
This is puzzling, since the enhanced radiation fields and enhanced
abundances found in proximate absorbers would be expected to lead to
stronger \nf. Indeed, \citet{FR09} report that the fraction of \cf\ systems
showing \nf\ rises from 11\% for intervening absorbers to 37\% for
proximate absorbers \citep[see also][]{Ha97, DO04}. 
One possible explanation for our apparent lack of a proximity effect
in DLA \nf\footnote{There is a proximity effect for DLAs in general,
  in that their incidence increases within several
  1\,000~\kms\ of the QSO \citep{El02, Ru06, Pr08b}.} 
is that the expected QSO sphere-of-influence (the region in which the
QSO radiation density exceeds the extragalactic background density)
extends over only $\sim$1\,500--2\,500~\kms\ at $z\!\approx\!2.5$, not
5\,000~\kms. In the current sample there are only a
handful of cases at $\la$2\,000~\kms\ from the QSO. 
Indeed, \citet{HW07} predict that absorbers with  $n_{\rm
H}\!\le\!0.1$~cm$^{-3}$  will be photo-evaporated by the
background quasar at $d\!\le\!1$~Mpc, or only $\la$250~\kms\ at
$z$=3 \citep[see also][]{PH09}. This picture is supported 
by the detection of strong \nf\ absorption in two highly-proximate
DLAs: one at 100$\pm$100~\kms\ from the quasar \citep{He09}, and
one extending over a few hundred \kms\ around $z_{\rm qso}$ \citep{Rx07}.  
Including these known systems in the sample raises the \nf\ detection rate
in proximate systems, but we still see no enhancement in the range
500--5\,000~\kms\ from the QSO.
This could be explained if the \nf\ largely arises in warm-hot rather than
photoionized plasma (see \S6.3), because the key
parameter governing the production of \nf\ in warm-hot plasma is
temperature, not UV radiation-field intensity.

On the other hand, \emph{foreground} quasars have been identified near two of
the twelve DLAs with \nf: in the field of \object{Q2243--6031},
\citet{Wk08} found two foreground QSOs at projected distances of 4.6
and 7.0~Mpc from the DLA at \zabs=2.33061; toward \object{Q0042-2930},
a foreground QSO identified by the 2dF QSO Redshift Survey
\citep[2QZ;][]{Cr04} lies at a projected of distance 1.4~Mpc from the
DLA at \zabs=1.80947. Accounting for these systems raises the
fraction of \nf\ DLAs that are associated with QSOs to $>$33\% (at
least four out of twelve). 

\subsection{The \nf\ b-value distribution}
\subsubsection{Narrow and broad components}
In the distribution of $b$(\nf), some $\approx$20\% of components are 
narrow ($b\!<\!10$~\kms) implying the absorbing plasma is at
log\,$T\!<\!4.92$. A potential explanation for these narrow \nf\
components is in collisionally-ionized plasma that has cooled out of
equilibrium. However, non-equilibrium cooling plasma models \citep{GS07}
can only produce \nf\ at log\,$T\!<\!5$ if the metallicity is close to
solar (Figure 8), and the metallicities in the DLAs under study are
generally below one tenth of solar. Thus photoionization is required as
the origin of the narrow components. Since \nf\ is difficult to
photoionize with starlight (even an O-type stellar spectrum is not
hard enough), and that EGB models require unreasonably large path
lengths, a nearby source of hard radiation, such as an AGN, is
required. 

The remaining $\approx$80\% of the \nf\ components have
$b\!>\!10$~\kms, with a tail extending to $b\!\ga\!30$~\kms\
(such a tail has also been noted in \cf\ and \sif\ in Paper I).
The median $b$(\nf) of 18.3~\kms\ implies 
log\,$T$=5.45 (assuming the line broadening is purely thermal), 
a temperature at which collisional ionization of \nf\ is
expected \citep{SD93, GS07}. Although we cannot rule out
  lower-temperature, photoionized solutions with 
a non-thermal contribution to the line widths,
\emph{the observed $b$-value distribution can be explained if a
substantial fraction of the \nf\ (and \cf\ and \sif) components arise
in warm-hot plasma.}
Note that the presence of such plasma in DLAs was already indicated by \os\
(Paper I), but the differing $b$-values of \nf\ than \os\ suggest
the plasma is multi-phase. We note that some DLAs contain both narrow
and broad \nf\ components (e.g. \object{Q0551-3637}; see Table 1),
suggesting that distinct regions of warm and hot plasma co-exist in
the same halos.

\subsubsection{Comparison: DLAs vs IGM vs Milky Way}
\citet{FR09} find a narrow median \nf\ $b$-value of only 6.0~\kms\ in
their study of \nf\ at high redshift, whereas our median
$b$(\nf) in DLAs is 18~\kms. The basic difference between the samples
is that the \citet{FR09} absorbers have \hi\ column densities in
the range log\,$N$(\hi)=12.5--16, whereas our sub-DLAs and DLAs all have
log\,$N$(\hi)$>$19.5. The differing $b$-values suggest a fundamental
difference between galactic and intergalactic \nf\ absorption, where
intergalactic \nf\ [identified by low accompanying $N$(\hi)]
is formed in warm, photoionized plasma, but the galactic \nf\
[identified by high $N$(\hi)] is largely formed in
warm-hot, collisionally-ionized plasma. 

In sight lines passing through the Galactic halo
(the Milky Way is, after all, a DLA),
weak \nf\ absorption is commonly observed in spectra of reasonable
S/N: \citet{IS04} report \nf\ detections in 32 of 34 sight lines surveyed.
The majority of Galactic \nf\ components show $b$-values between 30 and
80~\kms\ \citep{SS92, Sa95, Sa97, Se99, Fo03}. However, a handful of
interstellar \nf\ components with narrower $b$-values have been found, 
in sight lines passing through supershells \citep{Se97, Sa01} and the
Local Interstellar Medium \citep{WL05}. Thus \nf\ behaves similarly in
high-$z$ DLAs as in the Milky Way, with each case showing a majority
of broad components and a minority of narrow components. 
Another possible comparison is with GRB host galaxies
\citep[which are usually DLAs as well;][]{Ja06}.
\citet{Fo08b} find a median $b$-value of 18~\kms\ for the \nf\
components at the GRB redshift in a sample of seven $z\!>\!2$
afterglow spectra observed at high resolution, identical to the median
$b$(\nf) measured here in DLAs. 
Although the \nf\ profiles in GRB environments may be 
influenced by the strong flux of ionizing radiation from the burst
\citep{Pr08c}, this similarity in $b$-values suggests that GRB host
galaxies contain a similar phase of warm-hot interstellar plasma as do 
the DLA host galaxies under study here.

\subsection{Collisional ionization models}
To investigate the viability of collisional ionization (CI) as an origin
mechanism, we consider the CI models of \citet{GS07}, focusing on
three high-ion column density ratios: \nf/\sif, \nf/\cf, and \nf/\os.
For each of these three diagnostics, our approach is to: 
(1) derive the ratios expected in the models as a function of 
temperature\footnote{The \citet{GS07} ionization fractions are tabulated at 
  http://wise-obs.tau.ac.il/$\sim$orlyg/cooling/.}; 
(2) select only those temperatures at which both ionization fractions
(e.g. \nf/N and \cf/C for the case of the \nf/\cf\ ratio) 
are higher than a threshold of 0.003 (otherwise the high
ions would not be observed); (3) apply a correction 
for the relative elemental abundance ratios appropriate for DLAs
\citep[following the procedure described in][]{Fo04};
(4) compare these predictions to the observed ratios (see Figure 10).
We analyze three CI models: the equilibrium case
(CIE) and the non-equilibrium isobaric cases at [Z/H]=$-$1 and $-$2. 
We apply a correction [N/Si]=[N/O]=$-$1 (an average value for our
sample; see Table 2), and [N/C]=$-$0.5, which derives from combining
[N/O]=$-$1 with the finding that [C/O]$\approx-0.5$ in metal-poor halo
stars \citep{Ak04} and in the $z\!\approx\!2$ IGM \citep{Ag08}.

\begin{figure}[!ht]
\resizebox{\hsize}{!}{\includegraphics{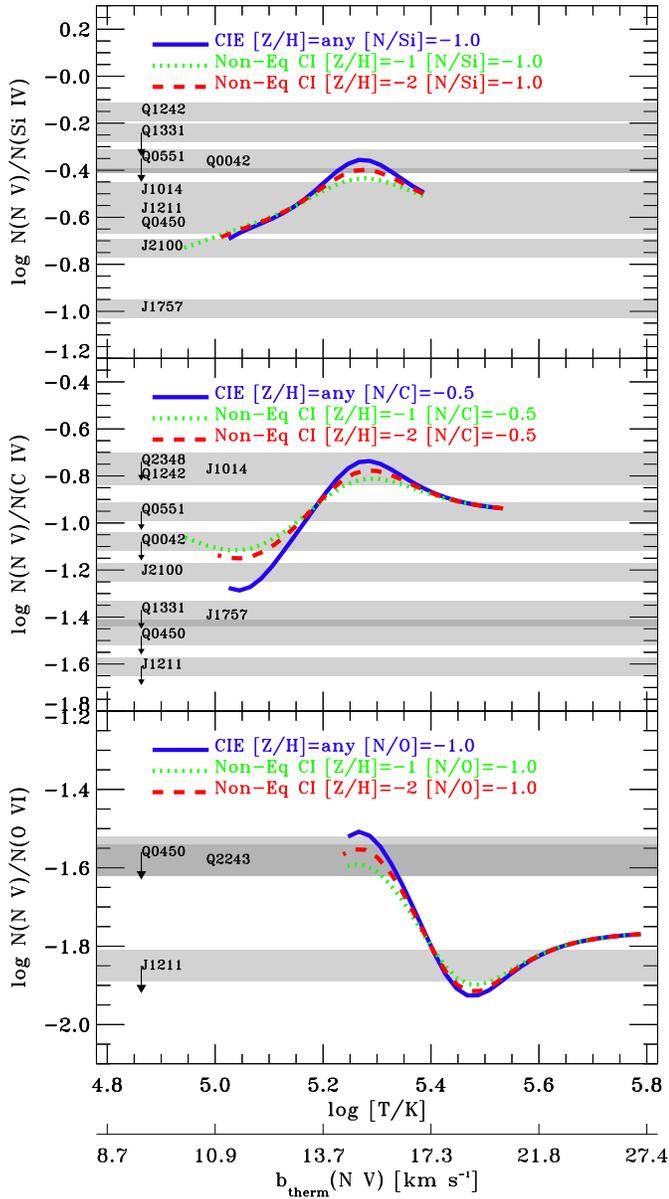}}
\caption{Comparison of the observed integrated high-ion column density
  ratios in DLAs (shaded regions) with the predictions of the
  collisional ionization models of \citet{GS07}. 
  Regions marked with arrows denote upper limits (due to saturation).
  Three models are shown in each panel, for the equilibrium case (CIE) and
  two non-equilibrium isobaric cases at [Z/H]=$-$1 and $-$2. 
  The model predictions have been adjusted for the expected [N/Si],
  [N/C], and [N/O] in DLAs (see text), and are only plotted for temperatures at
  which both ionization fractions $f$(\nf) and $f$(ion 2) are higher
  than a threshold of 0.003 (ion 2 is \sif, \cf, and \os, for the top,
  middle and bottom panels). The lower axis shows the corresponding
  \nf\ thermal line width.}
\end{figure}

Overall, the CI models are fairly successful at explaining the
integrated observed high-ion column density ratios (Fig 10). 
For each of the three ratios
studied (\nf/\sif, \nf/\cf, and \nf/\os), the model predictions
corrected for the DLA relative abundance patterns cover a similar
range as the observations. However, the range of temperature over
which each ratio has diagnostic power differs: 
\nf/\sif\ covers the range log\,$T$=5.0--5.4,
\nf/\cf\  covers log\,$T$=5.0--5.5, and 
\nf/\os\  covers log\,$T$=5.25--5.8.
Theoretically, a solution explaining all four ions is possible near
log\,$T$=5.3. However, such a solution predicts 
log[$N$(\nf)/$N$(\sif)]=$-$0.4,
log[$N$(\nf)/$N$(\cf)]=$-$0.8,
log[$N$(\nf)/$N$(\os)]=$-$1.6,
and no individual DLA in our sample matches all these values simultaneously.
Nonetheless, we cannot rule out a \emph{multi-phase} hot plasma model,
in which a superposition of absorbing regions exist at different
temperature, perhaps with the \nf\ and \cf\ components tracing regions
that have already cooled through the \os\ regime; indeed, a
multi-phase arrangement is strongly suggested by the $b$-value
distributions. To investigate such scenarios, detailed modelling of
individual components (rather than systems) is required, which is  
beyond the scope of this paper. Nonetheless, 
the fact that the \citet{GS07} CI models can readily reproduce (to
within a few tenths of a dex) the integrated high-ion ratios measured
in DLAs lends support to (but does not prove) the idea that the high
ions are formed in warm-hot plasma.

\subsection{Low-metallicity/strong-high-ion DLAs}
Among our sample are several DLAs with strong \nf\ absorption 
and very low elemental abundances, most notably the absorber at
$z_{\rm abs}$=2.2794 toward \object{Q2348--1444}, with
log\,$N$(\nf)=14.61$\pm$0.15 and [Si/H]=$-$1.92$\pm$0.02. 
Similarly, several DLAs with (fairly) strong \os\ but low O/H were
reported in Paper I, e.g. the systems at \zabs=2.6184 toward
\object{Q0913+0715}, with log\,$N$(\os)$>$14.43 and
[$\alpha$/H]=$-$2.59$\pm$0.10  \citep[see][]{Er06}, and at
\zabs=2.4560 toward \object{Q1409+0930}, with 
log\,$N$(\os)=14.27$\pm$0.03 and [$\alpha$/H]=$-$2.06$\pm$0.08.

These low-metallicity/strong-high-ion DLAs present interesting
challenges. If the dominant hot plasma production mechanism in
galactic halos is the shock-heating of infalling material
\citep[hot-mode accretion;][]{BD03, Ke05},  
then hot halos ($T\ga10^6$\,K) are only predicted to form around
galaxies with total mass $>10^{11.5}\,M_\odot$ \citep{DB06}; at lower
masses, the accretion occurs in cold mode.
If we combine this prediction with the suspected 
mass-metallicity relation in DLAs \citep[][though see Bouch\'e et
  al. 2008]{Le06, Pr08a}, then we only expect to see hot plasma in
the high metallicity (high mass) cases, but we cannot 
explain the low-metallicity/strong-high-ion DLAs.
On the other hand, it is equally difficult to explain these
unusual absorbers with an \emph{outflow}
model, since outflowing gas (driven by supernovae) should be metal-enriched.
We suggest that in these low-metallicity/strong-high-ion DLAs, there
may be abundance inhomogeneities between the neutral and
ionized regions, or the assumption that the \hi-absorbing
gas is neutral may be false. Detailed ionization modelling is required
to investigate their origin.

\section{Summary}
By combining datasets from VLT/UVES, Keck/HIRES, and
Keck/ESI, we have formed a large sample of high- and medium-resolution
quasar spectra, and surveyed \nf\ absorption in
91 DLAs and 18 sub-DLAs at redshifts between 1.67 and 
4.28 with unblended data at the wavelength of \nf.
The motivation for this work was to search for further evidence for
warm-hot plasma in high-redshift protogalactic structures, following the recent
detection of DLA \os. Our study has produced the following results.

\begin{enumerate}

\item 
Among DLAs, we find eight secure \nf\ detections 
(verified by both members of the doublet), four marginal \nf\ 
detections (seen in one member of the doublet only, but with a
consistent profile as the accompanying absorption in \cf), and 79 
non-detections. The detection rate of \nf\ in DLAs is therefore 9$^{+4}_{-3}$\%
(secure detections only) or 13$^{+5}_{-4}$\% (marginal detections included).
Among sub-DLAs, we find two secure \nf\ detection and 16 
non-detections at a similar detection rate of 11$^{+15}_{-7}$\%. 

\item
\nf\ is more likely to be detected in DLAs with high neutral-phase
metallicity. This is shown by significantly different (96.7\% 
probability) metallicity distributions of the DLAs with and without \nf\
detections. For the sub-sample of
absorbers with nitrogen abundance measurements, we find the \nf\
detection probability increases by a factor of $\approx$4 
above [N/H]=$-$2.3.
However, among the \nf\ detections, the \nf\ column density does
\emph{not} correlate with [N/H] or with [N/$\alpha$].
There is no significant trend for the \nf\
detection probability to depend on \hi\ column density. 
No \nf\ detections are found in DLAs at $z\!>\!3$, even though the
sample extends to $z\!\approx\!4$. 

\item
Four of the DLAs and one of the sub-DLAs with \nf\ detections also
show \os\ absorption, and all \ndet\ show \cf\ and \sif. 
The \nf\ column density correlates with both the \cf\ column density
and the \cf\ line width. The latter result, that \nf\ is
preferentially formed in the absorbers with the most extended
high-ion kinematics, suggests a link to star formation, since
absorbers with extended \cf\ velocity profiles can be explained by
supernova-driven winds.

\item
We investigate the $b$-value distributions of the \sif, \cf, and \nf\
components fit to the \ndet\ DLAs and sub-DLAs with \nf\ detections.
The median $b$-values for \sif, \cf, and \nf\ are 13.1, 16.3, and 
18.6~\kms, respectively. The \nf\ and \cf\ component distributions are
similar, suggesting these two ions largely arise in the same phase.
This is supported by optical depth comparisons of \cf\ and \nf\ (where
possible), which show that the two ions share similar profiles.
However, \os\ behaves differently, with a median $b$-value of
25.0~\kms\ (Paper I). This confirms that the plasma in DLAs and
sub-DLAs is multi-phase.

\item
Photoionization is implied for the origin of the narrow \nf\
components (which number $\approx$20\% of the total) with $b\!<\!10$~\kms\ and
log\,$T\!<\!4.92$. This is because neither non-equilibrium nor
equilibrium collisional models can produce \nf\ at these temperatures
given the sub-solar metallicities of the absorbers under study.
However, photoionization by the EGB is ruled out since the required
path lengths are a factor of 3--10 too large to be consistent with the
observed line widths, so local sources of ionizing radiation are needed.

\item
For the broad \nf\ (and \cf) components ($\approx$80\% of the total),
we cannot rule out photoionization by local sources of
hard radiation (e.g. foreground AGN), but we favor collisional
ionization in warm-hot plasma because:\\ 
(a) the median \nf\ line width in DLAs $b$=18~\kms\ is three times
broader than the median reported by \citet{FR09} in the (photoionized) IGM;\\ 
(b) collisional ionization models for plasma at log\,$T$=5.2--5.4 can 
reproduce (when corrected for the non-solar elemental
abundance ratios present in DLAs) the observed integrated high-ion
column density ratios in the majority of DLAs in our sample.\\ 
(c) there is no evidence for an enhanced \nf\ detection rate in
DLAs between 500 and 5\,000~\kms\ from the QSO,
as would be expected if the \nf\ was predominantly photoionized.

\item
We show that in a suite of collisional and photoionization models
covering a wide range of parameter space, the \nf\ ionization fraction
$f$(\nf) is always $<$0.25. Assuming that in each DLA, the
neutral-phase nitrogen abundance [N/H]=[\none/\hi] also applies to the
\nf\ phase, we derive lower limits on the total ionized hydrogen
column density in each absorber. If this assumption 
is correct, then in half our sample of DLAs with \nf, and in both
sub-DLAs with \nf, \emph{there is more mass in the \nf-phase alone
than in the neutral gas.} 
Taken together with the plasma's multi-phase nature, this implies
that even in those astrophysical regions where neutral gas accumulates
(presumably dark matter-dominated potential wells), warm-hot
plasma plays a significant role in the local baryon and metal budgets.

\end{enumerate}

\small {\bf Acknowledgments}. 
We thank Paolo Molaro for many useful comments, Bart
Wakker for advice on CLOUDY, and Gabor Worseck for
mentioning the role of foreground quasars.
AJF acknowledges support from an ESO Fellowship.
JXP and AMW are supported by NSF grant AST-0709235.
PP and RS acknowledge support from the Indo-French Centre
for the Promotion of Advanced Research (Centre Franco-Indien pour la
Promotion de la Recherche Avanc\'ee) under contract No. 3004-3.

\appendix
\section{Absorption-line measurements} 

\tiny
\begin{longtable}{lcccc ccccc cc}
\caption{DLA \nf\ measurements}\\
\hline\hline
QSO$^0$ & Ins.$^1$ & $z_{\rm em}$ & $z_{\rm abs}$$^2$ & log $N(\hi)$ & [Z/H]$^3$ & Z & Line$^4$ & For.?$^5$ & $v_-$, $v_+$ & $W_{\lambda}$(\nf)$^6$ & log\,$N(\nf)^7$\\
    &  & &  & ($N$ in cm$^{-2}$) & & & & & (\kms) & (m\AA) & ($N$ in cm$^{-2}$) \\
\hline
\endfirsthead
\caption{DLA \nf\ measurements (continued)}\\
\hline\hline
QSO$^0$ & Ins.$^1$ & $z_{\rm em}$ & $z_{\rm abs}$$^2$ & log $N(\hi)$ & [Z/H]$^3$ & Z & Line$^4$ & For.?$^5$ & $v_-$, $v_+$ & $W_{\lambda}$(\nf)$^6$ & log\,$N(\nf)^7$\\
    &  & &  & ($N$ in cm$^{-2}$) & & & & & (\kms) & (m\AA) & ($N$ in cm$^{-2}$) \\
\hline
\endhead
\hline
\endfoot
{\small {\bf Secure ID}}\\
    J1211+0422 & H & 2.54 & 2.37655 & 20.80$\pm$0.10 &      $-$1.41$\pm$0.08 & Si & 1238 & yes &       $-$130,     25 &                 76.7$\pm$3.6 &     13.63$\pm$0.22 \\
   Q0450--1310 & U & 2.25 & 2.06655 & 20.50$\pm$0.07 &      $-$1.45$\pm$0.05 & Si & 1238 & yes &       $-$ 30,    115 &                 65.6$\pm$2.3 &     13.56$\pm$0.02 \\
   Q0528--2505 & U & 2.77 & 2.81115 & 21.35$\pm$0.07 &      $-$0.90$\pm$0.11 & Si & 1238 &  no &       $-$115,    100 &                171.0$\pm$2.2 &     14.01$\pm$0.05 \\
   Q0551--3637 & U & 2.32 & 1.96221 & 20.70$\pm$0.08 &      $-$0.44$\pm$0.06 & Si & 1238 & yes &       $-$220,    100 &                319.9$\pm$9.7 &     14.28$\pm$0.07 \\
    Q1242+0006 & U & 2.08 & 1.82452 & 20.45$\pm$0.10 &      $-$1.04$\pm$0.07 & Si & 1238 & yes &       $-$865, $-$700 &                167.0$\pm$7.3 &     14.05$\pm$0.02 \\
    Q1331+1704 & U & 2.08 & 1.77635 & 21.15$\pm$0.07 &      $-$1.40$\pm$0.05 & Si & 1238 & yes &       $-$200,    200 &                224.7$\pm$3.1 &     14.08$\pm$0.06 \\
   Q2243--6031 & U & 3.01 & 2.33061 & 20.65$\pm$0.05 &      $-$0.85$\pm$0.05 & Si & 1238 & yes &       $-$380, $-$350 &        33.6$\pm$2.0$\dagger$ &     13.52$\pm$0.01 \\
   Q2348--1444 & U & 2.94 & 2.27940 & 20.56$\pm$0.08 &      $-$1.92$\pm$0.02 & Si & 1242 & yes &       $-$145,     25 &       214.3$\pm$2.2$\dagger$ &     14.61$\pm$0.15 \\
\hline
{\small {\bf Marginal ID}}\\
    J1014+4300 & H & 3.10 & 2.95880 & 20.50$\pm$0.02 &      $-$0.71$\pm$0.14 & Si & 1242 & yes &       $-$ 60,     25 &                 16.5$\pm$2.3 &     13.20$\pm$0.04 \\
    J1757+7539 & H & 3.05 & 2.62530 & 20.76$\pm$0.02 &      $-$0.79$\pm$0.01 & Si & 1242 & yes &       $-$ 80, $-$ 55 &                 14.8$\pm$2.2 &     13.20$\pm$0.03 \\
   J2100--0641 & E & 3.14 & 3.09240 & 21.05$\pm$0.15 &      $-$0.72$\pm$0.01 & Si & 1242 &  no &       $-$ 75,    100 &                 52.7$\pm$5.7 &     13.62$\pm$0.05 \\
   Q0042--2930 & U & 2.39 & 1.80947 & 20.40$\pm$0.10 &      $-$1.25$\pm$0.15 & Si & 1242 & yes &       $-$105, $-$ 75 &                 32.2$\pm$4.3 &     13.59$\pm$0.04 \\
\hline
{\small {\bf Non-detection}}\\
  HE0251--5550 & U & 2.37 & 2.33972 & 20.70$\pm$0.08 &      $-$1.41$\pm$0.08 &  S & 1238 &  no &       $-$ 50,     50 &                     $<$  7.9 &           $<$12.57 \\
  HE0322--3213 & U & 3.30 & 2.24338 & 20.50$\pm$0.10 &      $-$0.74$\pm$0.10 &  S & 1242 & yes &       $-$ 50,     50 &                     $<$ 88.3 &           $<$13.83 \\
  HE0414--2850 & U & 2.09 & 1.71903 & 21.20$\pm$0.10 &      $-$0.75$\pm$0.10 & Zn & 1242 & yes &       $-$ 50,     50 &                     $<$ 18.9 &           $<$13.25 \\
  HE2225--4025 & U & 2.41 & 1.96532 & 20.65$\pm$0.10 &      $-$2.26$\pm$0.10 & Si & 1242 & yes &       $-$ 50,     50 &                     $<$ 12.1 &           $<$13.05 \\
  HE2316--1012 & U & 1.95 & 1.92990 & 21.35$\pm$0.10 &      $-$1.57$\pm$0.12 & Zn & 1238 &  no &       $-$ 50,     50 &                     $<$ 17.0 &           $<$12.90 \\
    J0255+0008 & H & 3.97 & 3.91460 & 21.30$\pm$0.05 &      $-$1.78$\pm$0.01 &  S & 1242 &  no &       $-$ 20,     80 &                     $<$ 14.7 &           $<$13.14 \\
   J0339--0133 & H & 3.20 & 3.06210 & 21.20$\pm$0.10 &      $-$1.54$\pm$0.01 &  S & 1238 & yes &       $-$ 70,     70 &                     $<$ 27.3 &           $<$12.85 \\
    J0747+2739 & E & 4.11 & 3.42330 & 20.85$\pm$0.05 &      $-$1.66$\pm$0.21 & Si & 1238 & yes &       $-$180,     30 &                     $<$ 26.6 &           $<$13.10 \\
    J0812+3208 & H & 2.70 & 2.62630 & 21.35$\pm$0.10 &      $-$0.93$\pm$0.05 & Si & 1242 & yes &       $-$ 80,    210 &                     $<$  9.1 &           $<$12.93 \\
    J0836+1046 & H & 2.70 & 2.46530 & 20.58$\pm$0.10 &      $-$1.15$\pm$0.05 & Si & 1238 & yes &       $-$100,    200 &                     $<$ 86.2 &           $<$13.61 \\
    J0900+4204 & H & 3.29 & 3.24580 & 20.30$\pm$0.10 &      $-$0.85$\pm$0.02 &  S & 1238 &  no &       $-$ 15,     60 &                     $<$  6.2 &           $<$12.47 \\
    J0929+2825 & H & 3.40 & 3.26270 & 21.10$\pm$0.00 &      $-$1.62$\pm$0.01 & Si & 1242 & yes &       $-$ 50,    100 &                     $<$  7.0 &           $<$12.82 \\
    J0930+2803 & H & 3.42 & 3.23530 & 20.30$\pm$0.10 &      $-$1.97$\pm$0.02 & Si & 1242 & yes &       $-$ 50,      0 &                     $<$  8.4 &           $<$12.89 \\
    J1021+3001 & H & 3.12 & 2.94890 & 20.70$\pm$0.10 &      $-$1.94$\pm$0.02 & Si & 1238 & yes &       $-$ 25,     70 &                     $<$ 11.7 &           $<$12.74 \\
    J1131+6044 & H & 2.92 & 2.87600 & 20.50$\pm$0.15 &      $-$2.13$\pm$0.15 & Si & 1238 &  no &       $-$100,     50 &                     $<$ 15.2 &           $<$12.85 \\
    J1135+2227 & H & 2.88 & 2.78350 & 21.00$\pm$0.07 &      $-$2.08$\pm$0.14 & Si & 1242 & yes &       $-$ 50,     50 &                     $<$ 16.6 &           $<$12.92 \\
    J1310+5424 & H & 1.93 & 1.80070 & 21.45$\pm$0.15 &      $-$0.45$\pm$0.15 & Zn & 1242 & yes &       $-$100,    100 &                     $<$ 81.9 &           $<$13.88 \\
   J1357--1744 & H & 3.15 & 2.77990 & 20.30$\pm$0.15 &      $-$1.86$\pm$0.05 & Si & 1238 & yes &       $-$ 50,      0 &                     $<$ 53.6 &           $<$13.26 \\
    J1410+5111 & H & 3.21 & 2.96420 & 20.85$\pm$0.20 &      $-$1.96$\pm$0.15 & Si & 1238 & yes &           75,    200 &                     $<$ 21.7 &           $<$13.01 \\
    J1435+5359 & H & 2.64 & 2.34270 & 21.05$\pm$0.10 &      $-$1.48$\pm$0.10 & Si & 1238 & yes &       $-$ 50,     70 &                     $<$  9.5 &           $<$12.65 \\
    J1506+5220 & H & 4.18 & 3.22440 & 20.67$\pm$0.07 &      $-$2.33$\pm$0.02 & Si & 1238 & yes &       $-$ 50,     30 &                     $<$ 21.3 &           $<$13.00 \\
   J1558--0031 & H & 2.83 & 2.70260 & 20.67$\pm$0.05 &      $-$1.99$\pm$0.01 & Si & 1238 & yes &       $-$ 20,     20 &                     $<$ 38.4 &           $<$13.16 \\
   J2036--0552 & H & 2.58 & 2.28050 & 21.20$\pm$0.15 &      $-$1.71$\pm$0.17 & Si & 1238 & yes &       $-$ 30,     50 &                     $<$ 13.1 &           $<$12.79 \\
    J2241+1352 & E & 4.44 & 4.28240 & 21.15$\pm$0.10 &      $-$1.76$\pm$0.03 &  S & 1242 & yes &       $-$ 15,    150 &                     $<$ 64.8 &           $<$13.53 \\
    J2323+2758 & E & 4.18 & 3.68450 & 20.95$\pm$0.10 &      $-$2.59$\pm$0.03 & Si & 1238 & yes &       $-$ 50,     50 &                     $<$ 79.1 &           $<$13.35 \\
   J2340--0053 & H & 2.09 & 2.05450 & 20.35$\pm$0.15 &      $-$0.74$\pm$0.04 & Si & 1238 &  no &       $-$ 50,     50 &                     $<$ 22.5 &           $<$12.87 \\
    J2346+1245 & H & 2.79 & 2.53790 & 20.36$\pm$0.10 &      $-$1.74$\pm$0.01 & Si & 1242 & yes &       $-$ 50,     10 &                     $<$ 32.5 &           $<$13.32 \\
   Q0010--0012 & U & 2.15 & 2.02478 & 20.95$\pm$0.10 &      $-$1.43$\pm$0.11 & Zn & 1238 & yes &       $-$ 50,     50 &                     $<$ 21.9 &           $<$13.01 \\
   Q0039--3354 & U & 2.48 & 2.22400 & 20.60$\pm$0.10 &      $-$1.31$\pm$0.12 & Si & 1242 & yes &       $-$ 50,    130 &                     $<$ 47.8 &           $<$13.40 \\
   Q0042--2930 & U & 2.39 & 1.93560 & 20.50$\pm$0.10 &      $-$1.27$\pm$0.10 & Si & 1242 & yes &       $-$ 50,     50 &                     $<$ 32.4 &           $<$13.48 \\
   Q0049--2820 & U & 2.26 & 2.07125 & 20.45$\pm$0.10 &      $-$1.31$\pm$0.12 & Si & 1242 & yes &       $-$ 20,     50 &                     $<$ 11.0 &           $<$13.01 \\
    Q0100+1300 & U & 2.69 & 2.30904 & 21.35$\pm$0.08 &      $-$1.46$\pm$0.01 &  S & 1242 & yes &       $-$ 50,     50 &                     $<$ 20.2 &           $<$13.08 \\
   Q0102--1902 & U & 3.04 & 2.36962 & 21.00$\pm$0.08 &      $-$1.90$\pm$0.08 &  S & 1242 & yes &       $-$ 50,     50 &                     $<$ 10.3 &           $<$12.98 \\
    Q0112+0259 & U & 2.81 & 2.42310 & 20.90$\pm$0.10 &      $-$1.31$\pm$0.11 &  S & 1238 & yes &       $-$ 50,    100 &                     $<$ 13.1 &           $<$12.79 \\
   Q0112--3030 & U & 2.99 & 2.41850 & 20.50$\pm$0.08 &      $-$2.42$\pm$0.08 & Si & 1238 & yes &       $-$ 30,     30 &                     $<$  9.1 &           $<$12.63 \\
   Q0135--2722 & U & 3.21 & 2.10739 & 20.30$\pm$0.15 &      $-$1.12$\pm$0.16 &  S & 1238 & yes &            0,    120 &                     $<$ 24.0 &           $<$13.05 \\
   Q0135--2722 & U & 3.21 & 2.80003 & 21.00$\pm$0.10 &      $-$1.40$\pm$0.10 &  S & 1242 & yes &       $-$100,     50 &                     $<$ 73.6 &           $<$13.73 \\
   Q0254--4025 & U & 2.28 & 2.04607 & 20.45$\pm$0.08 &      $-$1.55$\pm$0.09 &  S & 1238 & yes &       $-$ 50,     50 &                     $<$ 12.7 &           $<$12.77 \\
   Q0300--3152 & U & 2.37 & 2.17905 & 20.80$\pm$0.10 &      $-$1.80$\pm$0.10 &  S & 1238 & yes &       $-$ 50,     50 &                     $<$ 50.0 &           $<$13.21 \\
   Q0405--4418 & U & 3.02 & 1.91270 & 20.80$\pm$0.10 &      $-$1.03$\pm$0.10 & Zn & 1238 & yes &       $-$100,     40 &                     $<$ 68.9 &           $<$13.45 \\
   Q0405--4418 & U & 3.02 & 2.54990 & 21.15$\pm$0.15 &      $-$1.36$\pm$0.16 & Zn & 1242 & yes &            0,    150 &                     $<$ 45.8 &           $<$13.52 \\
   Q0405--4418 & U & 3.02 & 2.59475 & 21.05$\pm$0.10 &      $-$1.12$\pm$0.10 & Zn & 1238 & yes &       $-$ 50,     70 &                     $<$ 58.1 &           $<$13.38 \\
   Q0405--4418 & U & 3.02 & 2.62140 & 20.45$\pm$0.10 &      $-$2.04$\pm$0.10 & Si & 1242 & yes &           70,    150 &                     $<$ 27.9 &           $<$13.30 \\
   Q0421--2624 & U & 2.28 & 2.15680 & 20.65$\pm$0.10 &      $-$1.86$\pm$0.10 & Si & 1238 & yes &            0,    230 &                     $<$ 24.0 &           $<$13.05 \\
   Q0425--5214 & U & 2.25 & 2.22430 & 20.30$\pm$0.10 &      $-$1.43$\pm$0.11 &  S & 1238 &  no &       $-$ 75,     50 &                     $<$  9.9 &           $<$12.67 \\
   Q0432--4401 & U & 2.65 & 2.30197 & 20.95$\pm$0.10 &      $-$1.23$\pm$0.13 &  S & 1238 & yes &       $-$150, $-$ 70 &                     $<$ 12.7 &           $<$12.78 \\
   Q0458--0203 & U & 2.29 & 2.03955 & 21.70$\pm$0.10 &      $-$1.22$\pm$0.10 & Zn & 1242 & yes &       $-$150,     50 &                     $<$ 79.3 &           $<$13.67 \\
   Q0642--5038 & U & 3.09 & 2.65860 & 20.95$\pm$0.08 &      $-$1.05$\pm$0.09 & Zn & 1242 & yes &       $-$100,    100 &                     $<$ 22.7 &           $<$13.33 \\
    Q0841+1256 & U & 2.50 & 1.86388 & 21.00$\pm$0.10 &      $-$1.51$\pm$0.11 &  S & 1238 & yes &       $-$100,     10 &                     $<$ 38.0 &           $<$13.06 \\
    Q0841+1256 & U & 2.50 & 2.37452 & 21.05$\pm$0.10 &      $-$1.27$\pm$0.02 & Si & 1238 & yes &       $-$ 50,     50 &                     $<$ 14.1 &           $<$12.82 \\
    Q0841+1256 & U & 2.50 & 2.47621 & 20.80$\pm$0.10 &      $-$1.39$\pm$0.03 &  S & 1238 &  no &       $-$ 50,     50 &                     $<$  9.2 &           $<$12.64 \\
   Q0933--3319 & U & 2.68 & 2.68227 & 20.50$\pm$0.10 &      $-$1.27$\pm$0.14 & Si & 1238 &  no &       $-$ 50,     50 &                     $<$ 27.1 &           $<$13.10 \\
   Q1036--2257 & U & 3.13 & 2.77739 & 20.93$\pm$0.05 &      $-$1.36$\pm$0.05 &  S & 1238 & yes &       $-$ 15,    150 &                     $<$ 21.8 &           $<$12.79 \\
   Q1055--3008 & U & 1.90 & 1.90350 & 21.54$\pm$0.10 &      $-$1.30$\pm$0.03 & Zn & 1238 &  no &       $-$ 50,     50 &                     $<$ 31.6 &           $<$13.17 \\
   Q1108--0747 & U & 3.92 & 3.60760 & 20.37$\pm$0.07 &      $-$1.59$\pm$0.07 & Si & 1238 & yes &       $-$130, $-$ 25 &                     $<$ 12.2 &           $<$12.76 \\
   Q1111--1517 & U & 3.37 & 3.26548 & 21.30$\pm$0.05 &      $-$1.65$\pm$0.11 & Zn & 1242 & yes &           70,    170 &                     $<$  6.1 &           $<$12.76 \\
    Q1157+0128 & U & 1.99 & 1.94375 & 21.80$\pm$0.10 &      $-$1.44$\pm$0.10 & Zn & 1238 &  no &       $-$ 75,     50 &                     $<$ 12.8 &           $<$12.78 \\
    Q1203+0218 & U & 2.13 & 1.74735 & 20.40$\pm$0.10 &      $-$0.97$\pm$0.11 & Zn & 1242 & yes &       $-$100,     15 &                     $<$ 72.7 &           $<$13.64 \\
    Q1210+1731 & U & 2.54 & 1.89177 & 20.70$\pm$0.08 &      $-$0.93$\pm$0.08 &  S & 1242 & yes &       $-$ 50,     50 &                     $<$ 71.3 &           $<$13.77 \\
    Q1223+1753 & U & 2.94 & 2.46608 & 21.40$\pm$0.10 &      $-$1.63$\pm$0.10 & Zn & 1242 & yes &       $-$ 85,    150 &                     $<$ 54.6 &           $<$13.55 \\
    Q1337+1121 & U & 2.92 & 2.79583 & 21.00$\pm$0.08 &      $-$1.86$\pm$0.09 & Si & 1238 & yes &       $-$ 80,     25 &                     $<$ 13.8 &           $<$12.81 \\
   Q1354--1046 & U & 3.01 & 2.96680 & 20.80$\pm$0.10 &      $-$1.39$\pm$0.10 & Si & 1238 &  no &       $-$ 50,     50 &                     $<$  7.7 &           $<$12.56 \\
    Q1409+0930 & U & 2.85 & 2.01882 & 20.65$\pm$0.10 &      $-$1.62$\pm$0.16 & Zn & 1238 & yes &       $-$ 50,     50 &                     $<$  8.5 &           $<$12.60 \\
    Q1409+0930 & U & 2.85 & 2.45595 & 20.53$\pm$0.08 &      $-$2.06$\pm$0.08 & Si & 1238 & yes &       $-$ 50,     50 &                     $<$ 29.0 &           $<$12.96 \\
   Q2059--3604 & U & 3.09 & 3.08291 & 20.98$\pm$0.08 &      $-$1.77$\pm$0.09 &  S & 1242 &  no &       $-$ 25,    100 &                     $<$ 16.0 &           $<$13.17 \\
   Q2138--4427 & U & 3.17 & 2.85235 & 20.98$\pm$0.05 &      $-$1.74$\pm$0.05 & Zn & 1242 & yes &       $-$270,    120 &                     $<$ 10.7 &           $<$13.00 \\
   Q2206--1958 & U & 2.56 & 2.07622 & 20.44$\pm$0.05 &      $-$2.32$\pm$0.05 & Si & 1242 & yes &       $-$ 25,     50 &                     $<$ 16.7 &           $<$12.99 \\
   Q2222--3939 & U & 2.18 & 2.15387 & 20.85$\pm$0.10 &      $-$1.97$\pm$0.10 &  S & 1242 &  no &       $-$ 25,     25 &                     $<$  8.1 &           $<$12.88 \\
   Q2228--3954 & U & 2.21 & 2.09437 & 21.20$\pm$0.10 &      $-$1.36$\pm$0.12 & Zn & 1242 & yes &       $-$ 35,     25 &                     $<$ 18.6 &           $<$13.24 \\
    Q2230+0232 & U & 2.15 & 1.86377 & 20.90$\pm$0.10 &      $-$0.81$\pm$0.10 &  S & 1238 & yes &       $-$ 50,     50 &                     $<$ 10.6 &           $<$12.70 \\
   Q2311--3721 & U & 2.48 & 2.18210 & 20.48$\pm$0.13 &              $<-$1.33 & Zn & 1242 & yes &       $-$100,     50 &                     $<$ 10.8 &           $<$13.00 \\
   Q2314--4057 & U & 2.45 & 1.85733 & 20.90$\pm$0.10 &      $-$1.02$\pm$0.14 & Zn & 1238 & yes &       $-$100,     50 &                     $<$ 25.0 &           $<$13.07 \\
   Q2332--0924 & U & 3.32 & 3.05725 & 20.50$\pm$0.07 &      $-$1.33$\pm$0.08 &  S & 1238 & yes &       $-$130,     50 &                     $<$ 11.4 &           $<$12.73 \\
    Q2341+0325 & U & 4.24 & 3.22030 & 21.35$\pm$0.07 &              $<-$1.63 & Zn & 1238 & yes &       $-$ 20,     70 &                     $<$ 31.9 &           $<$13.00 \\
    Q2342+3417 & U & 3.01 & 2.90910 & 21.10$\pm$0.10 &      $-$1.04$\pm$0.02 & Si & 1238 & yes &       $-$ 50,     50 &                     $<$ 38.7 &           $<$13.12 \\
    Q2343+1232 & U & 2.76 & 2.43127 & 20.40$\pm$0.07 &      $-$0.60$\pm$0.06 & Si & 1242 & yes &       $-$350, $-$ 50 &                     $<$ 17.3 &           $<$12.92 \\
   Q2348--0108 & U & 3.01 & 2.61473 & 21.30$\pm$0.08 &      $-$2.02$\pm$0.08 & Si & 1242 & yes &       $-$ 50,     50 &                     $<$ 10.5 &           $<$12.99 \\
   Q2359--0216 & U & 2.81 & 2.09508 & 20.65$\pm$0.10 &      $-$0.84$\pm$0.13 & Zn & 1242 & yes &       $-$180,    100 &                     $<$ 44.8 &           $<$13.62 \\
   Q2359--0216 & U & 2.81 & 2.15390 & 20.30$\pm$0.10 &      $-$1.62$\pm$0.10 & Si & 1238 & yes &       $-$150,     75 &                     $<$ 52.3 &           $<$13.39 \\
\end{longtable}
\noindent $^0$ Our adopted QSO names follow one of three conventions, each based on the object coordinates. The UVES quasars are named either HEhhmm+ddmm (Hamburg-ESO survey, J2000) or Qhhmm+ddmm (J1950); the HIRES and ESI quasars are named Jhhmm+ddmm (J2000). We retain these (non-standardized) names to facilitate comparison with previous work. Absorbers are listed alphabetically within each category.\\
\noindent $^1$ Instrument that provided the data: U=UVES, H=HIRES, E=ESI.\\
$^2$ Redshift defined by velocity of strongest low-ionization component.\\
$^3$ Metallicity on log scale relative to solar. Z=S, Si, or Zn as listed in adjacent column. Most [Z/H] and log\,$N$(\hi) measurements are from \citet{Le06}, \citet{Pr07}, and \citet{Wo08}. Several others were taken from \citet{Lu96}, \citet{EL01}, \citet{SP01}, \citet{Ak05}, \citet{Hu06}, \citet{Fo07b}, \citet{No08}, or Kaplan et al. (2009, in prep.).\\
$^4$ \nf\ line used for measurement. \\
$^5$ Is the \nf\ doublet in the \lya\ forest? If no, the system is within 5\,600~\kms\ of background QSO, so ``no" implies ``proximate".\\
$^6$ \nf\ rest-frame equivalent width, measured in the velocity range listed in the adjacent column. Upper limits are 3$\sigma$. In cases marked $\dagger$, the data are partly blended; the measurement listed here derives from the unblended velocity range, and the total $W_{\lambda}$ could be higher.\\
$^7$ For detections, $N_{\rm \nf}$ is the sum of the component column densities from VPFIT. For non-detections, 3$\sigma$ upper limits were calculated using $N_{\rm \nf}=1.13\times10^{17}W_{\lambda}^{\rm lim}/\lambda_0^2 f$, where $W_{\lambda}^{\rm lim}$ is the 3$\sigma$ upper limit on the equivalent width in m\AA, $\lambda_0$ is in \AA, and $N_{\rm \nf}$ is in cm$^{-2}$.
 The \nf\ doublet has $\lambda_0$=1238.821, 1242.804\,\AA\ and $f$=0.157, 0.07821 respectively \citep{Mo03}.\\
{\bf List of DLAs with blended \nf\ (QSO, $z_{\rm abs}$, instrument):}
HE0242--2917 2.55956 U; 
J0022--1505 3.43890 H; 
J0127--0045 3.72740 E; 
J0133+0400 3.69260 H; 
J0134+3307 3.76090 E; 
J0203+1134 3.38690 H; 
J0209+0517 3.66670 H; 
J0209+0517 3.86430 H; 
J0349--3810 3.02470 H; 
J0426--2202 2.98310 E; 
J0747+2739 3.90000 E; 
J0808+5215 3.11320 E; 
J0826+3148 2.91220 H; 
J0956+4734 3.40360 E; 
J0956+4734 3.89100 E; 
J0956+4734 4.24420 E; 
J1057+4555 3.31720 H; 
J1155+0530 3.32680 H; 
J1248+3110 3.69700 E; 
J1253--0228 2.78280 E; 
J1410+5111 2.93440 H; 
J1432+3940 3.27250 E; 
J1502+4803 2.56960 E; 
J1723+2243 3.69470 E; 
J2225+2040 3.11920 H; 
Q0013--0029 1.97295 U; 
Q0027--1836 2.40186 U; 
Q0058--2914 2.67140 U; 
Q0112--3030 2.70230 U; 
Q0130+0345 3.77373 U; 
Q0216+0803 1.76875 U; 
Q0438--4338 2.34736 U; 
Q0528--2505 2.14105 U; 
Q0913+0715 2.61840 U; 
Q0951--0450 3.85670 H; 
Q0951--0450 4.20287 U; 
Q0952--0115 4.02270 U; 
Q1117--1329 3.35037 U; 
Q1209+0919 2.58440 U; 
Q1228--1122 2.19289 U; 
Q1232+0815 2.33771 U; 
Q1354--1046 2.50090 U; 
Q1441+2737 4.22371 U; 
Q1451+1223 2.25465 U; 
Q1451+1223 2.46921 U; 
Q2138--4427 2.38279 U; 
Q2153+1344 3.31600 U; 
Q2206--1958 1.91998 U; 
Q2318--1107 1.98888 U; 
Q2348--0108 2.42630 U.

\tiny
\begin{longtable}{lcccc ccccc cc}
\caption{Sub-DLA \nf\ measurements}\\
\hline\hline
QSO$^0$ & Ins.$^1$ & $z_{\rm em}$ & $z_{\rm abs}$$^2$ & log $N(\hi)$ & [Z/H]$^3$ & Z & Line$^4$ & For.?$^5$ & $v_-$, $v_+$ & $W_{\lambda}$(\nf)$^6$ & log\,$N(\nf)^7$\\
    &  & &  & ($N$ in cm$^{-2}$) & & & & & (\kms) & (m\AA) & ($N$ in cm$^{-2}$) \\
\hline
\endfirsthead
\caption{Sub-DLA \nf\ measurements (continued)}\\
\hline\hline
QSO$^0$ & Ins.$^1$ & $z_{\rm em}$ & $z_{\rm abs}$$^2$ & log $N(\hi)$ & [Z/H]$^3$ & Z & Line$^4$ & For.?$^5$ & $v_-$, $v_+$ & $W_{\lambda}$(\nf)$^6$ & log\,$N(\nf)^7$\\
    &  & &  & ($N$ in cm$^{-2}$) & & & & & (\kms) & (m\AA) & ($N$ in cm$^{-2}$) \\
\hline
\endhead
\hline
\endfoot
{\small {\bf Detection}}\\
   Q0237--2321 & U & 2.22 & 1.67234 & 19.65$\pm$0.10 &      $-$0.26$\pm$0.10 & Si & 1238 & yes &       $-$ 50,    100 &                101.7$\pm$2.3 &     13.88$\pm$0.19 \\
   Q1037--2704 & U & 2.20 & 2.13906 & 19.70$\pm$0.05 &      $-$0.02$\pm$0.10 & Si & 1238 & yes &       $-$340, $-$280 &        67.1$\pm$2.0$\dagger$ &     13.77$\pm$0.01 \\
\hline
{\small {\bf Non-detection}}\\
  HE0057--4126 & U & 2.05 & 1.72055 & 20.13$\pm$0.07 &      $-$2.06$\pm$0.08 & Si & 1238 & yes &       $-$ 50,     50 &                     $<$ 18.6 &           $<$12.94 \\
  HE0438--1638 & U & 1.97 & 1.69420 & 19.85$\pm$0.10 &      $-$0.93$\pm$0.10 & Si & 1238 & yes &       $-$ 50,     50 &                     $<$ 30.9 &           $<$13.16 \\
    J0953+5230 & H & 1.87 & 1.76780 & 20.10$\pm$0.10 &      $-$0.05$\pm$0.05 & Si & 1242 & yes &       $-$100,    200 &                     $<$ 58.0 &           $<$13.51 \\
   Q0049--2820 & U & 2.26 & 1.88615 & 20.20$\pm$0.08 &      $-$1.03$\pm$0.09 &  S & 1238 & yes &       $-$ 50,     50 &                     $<$ 24.4 &           $<$13.06 \\
   Q0102--1902 & U & 3.04 & 2.92650 & 20.00$\pm$0.10 &      $-$1.50$\pm$0.10 & Si & 1238 & yes &       $-$ 50,     50 &                     $<$  7.0 &           $<$12.51 \\
   Q0331--4505 & U & 2.67 & 2.41120 & 20.15$\pm$0.07 &              $<-$1.41 & Zn & 1238 & yes &       $-$ 40,     50 &                     $<$ 12.3 &           $<$12.76 \\
   Q0331--4505 & U & 2.67 & 2.65618 & 19.82$\pm$0.05 &      $-$1.49$\pm$0.05 & Si & 1238 &  no &       $-$ 50,     50 &                     $<$ 11.8 &           $<$12.74 \\
   Q1101--2629 & U & 2.14 & 1.83890 & 19.35$\pm$0.04 &      $-$0.82$\pm$0.14 &  S & 1242 & yes &       $-$ 65,     50 &                     $<$ 16.2 &           $<$12.98 \\
   Q1220--1800 & U & 2.16 & 2.11285 & 20.12$\pm$0.07 &      $-$0.93$\pm$0.07 &  S & 1238 &  no &       $-$ 60,     60 &                     $<$ 11.3 &           $<$12.72 \\
    Q1337+1121 & U & 2.92 & 2.50792 & 20.12$\pm$0.05 &      $-$1.81$\pm$0.06 & Si & 1242 & yes &       $-$ 50,     50 &                     $<$ 25.1 &           $<$13.37 \\
   Q1340--1340 & U & 3.20 & 3.11835 & 20.05$\pm$0.08 &      $-$1.42$\pm$0.08 &  S & 1242 & yes &       $-$125,    250 &                     $<$  6.6 &           $<$12.79 \\
    Q1409+0930 & U & 2.85 & 2.66818 & 19.80$\pm$0.08 &      $-$1.41$\pm$0.09 &  S & 1242 & yes &       $-$200,      0 &                     $<$ 37.8 &           $<$13.43 \\
    Q1444+0126 & U & 2.21 & 2.08679 & 20.25$\pm$0.07 &      $-$0.80$\pm$0.09 & Zn & 1242 & yes &          100,    190 &                     $<$ 11.2 &           $<$13.02 \\
   Q2059--3604 & U & 3.09 & 2.50735 & 20.29$\pm$0.07 &      $-$1.85$\pm$0.20 &  S & 1238 & yes &       $-$ 50,     50 &                     $<$ 56.0 &           $<$13.26 \\
    Q2153+1344 & U & 4.26 & 4.21225 & 19.70$\pm$0.10 &      $-$1.81$\pm$0.10 & Si & 1242 &  no &       $-$ 50,     50 &                     $<$  7.3 &           $<$12.83 \\
   Q2314--4057 & U & 2.45 & 1.87519 & 20.10$\pm$0.20 &              $<-$1.19 & Zn & 1238 & yes &       $-$ 20,     50 &                     $<$ 11.0 &           $<$12.71 \\
\end{longtable}
\noindent $^0$ See footnote to Table A.1 for naming conventions.\\
\noindent $^1$ Instrument that provided the data: U=UVES, H=HIRES, E=ESI.\\
$^2$ Redshift defined by velocity of strongest low-ionization component.\\
$^3$ Metallicity on log scale relative to solar. Z=S, Si, or Zn as listed in adjacent column. No ionization corrections have been applied.\\
$^4$ \nf\ line used for measurement. \\
$^5$ Is the \nf\ doublet in the \lya\ forest? If no, the system is within 5\,600~\kms\ of background QSO, so ``no" implies ``proximate".\\
$^6$ \nf\ rest-frame equivalent width, measured in the velocity range listed in the adjacent column. Upper limits are 3$\sigma$. In cases marked $\dagger$, the data are partly blended; the measurement listed here derives from the unblended velocity range, and the total $W_{\lambda}$ could be higher.\\
$^7$ For detections, $N_{\rm \nf}$ is the sum of the component column densities from VPFIT. For non-detections, 3$\sigma$ upper limits were calculated using $N_{\rm \nf}=1.13\times10^{17}W_{\lambda}^{\rm lim}/\lambda_0^2 f$, where $W_{\lambda}^{\rm lim}$ is the 3$\sigma$ upper limit on the equivalent width in m\AA, $\lambda_0$ is in \AA, and $N_{\rm \nf}$ is in cm$^{-2}$.
 \\
{\bf List of sub-DLAs with blended \nf\ (QSO, $z_{\rm abs}$, instrument):}
Q0952--0115 3.47570 U; 
Q1108--0747 3.48179 U; 
Q2116--3537 1.99618 U; 
Q2153+1344 3.14198 U; 
Q2332--0924 2.28749 U.

\begin{figure*}
\includegraphics[width=18.5cm]{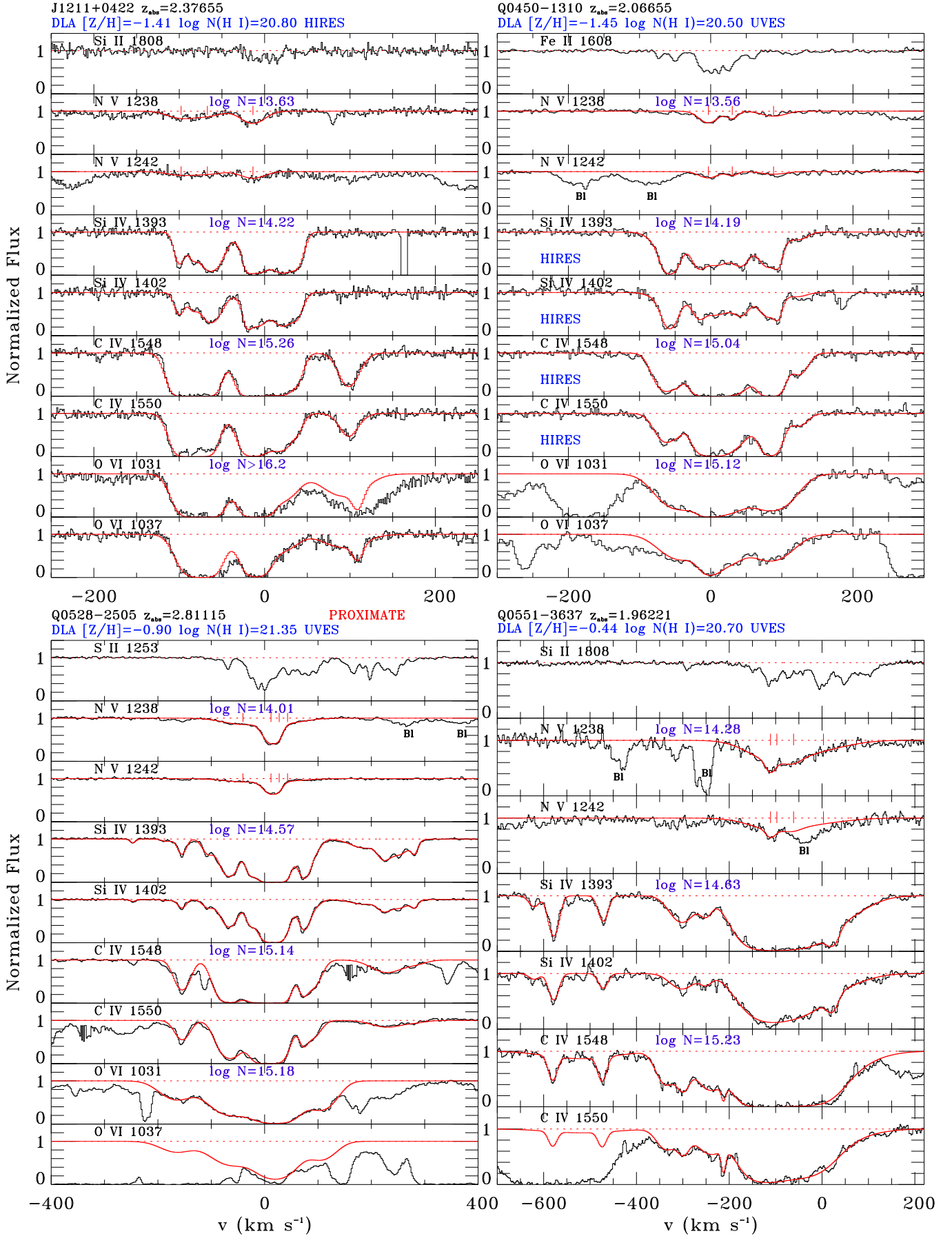}
\caption{Normalized high-ion absorption-line profiles for four DLAs
  with secure \nf\ detections. A low-ionization line is also shown.
  VPFIT Voigt-profile models are included as solid red lines.
  In the \nf\ panels, blends are indicated with the label `Bl', and
  tick marks show the component centers in our best-fit model.
  The velocity zero-point is defined by the DLA redshift.
  We annotate on the high-ion panels the total column density
  obtained by VPFIT.}
\end{figure*}
\begin{figure*}
\includegraphics[width=18.5cm]{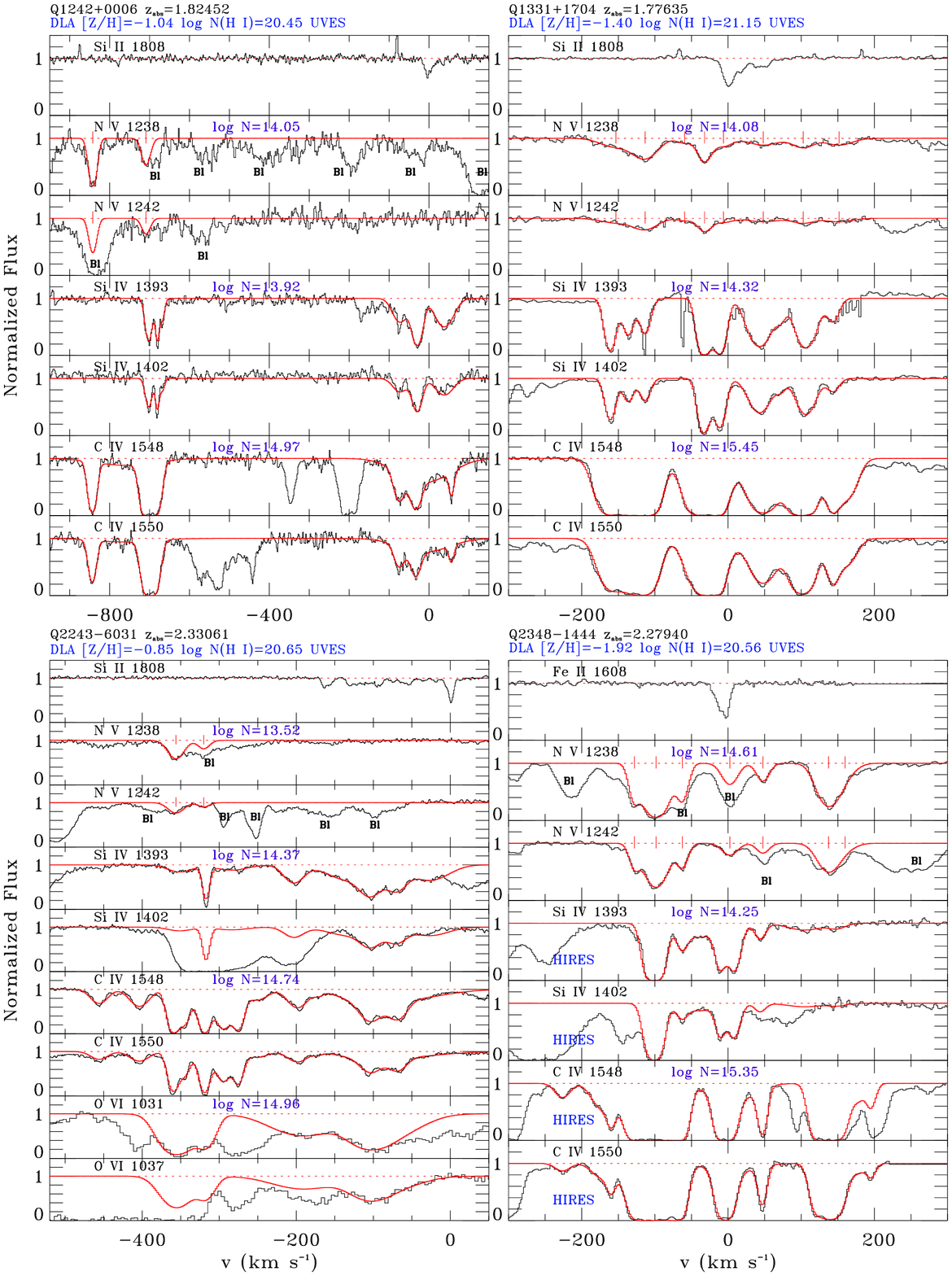}
\caption{Normalized high-ion absorption-line profiles for four DLAs
  with secure \nf\ detections. A low-ionization line is also shown.
  VPFIT Voigt-profile models are included as solid red lines.
  In the \nf\ panels, blends are indicated with the label `Bl', and
  tick marks show the component centers in our best-fit model.
  The velocity zero-point is defined by the DLA redshift.
  We annotate on the high-ion panels the total column density
  obtained by VPFIT.} 
\end{figure*}
\begin{figure*}
\includegraphics[width=18.5cm]{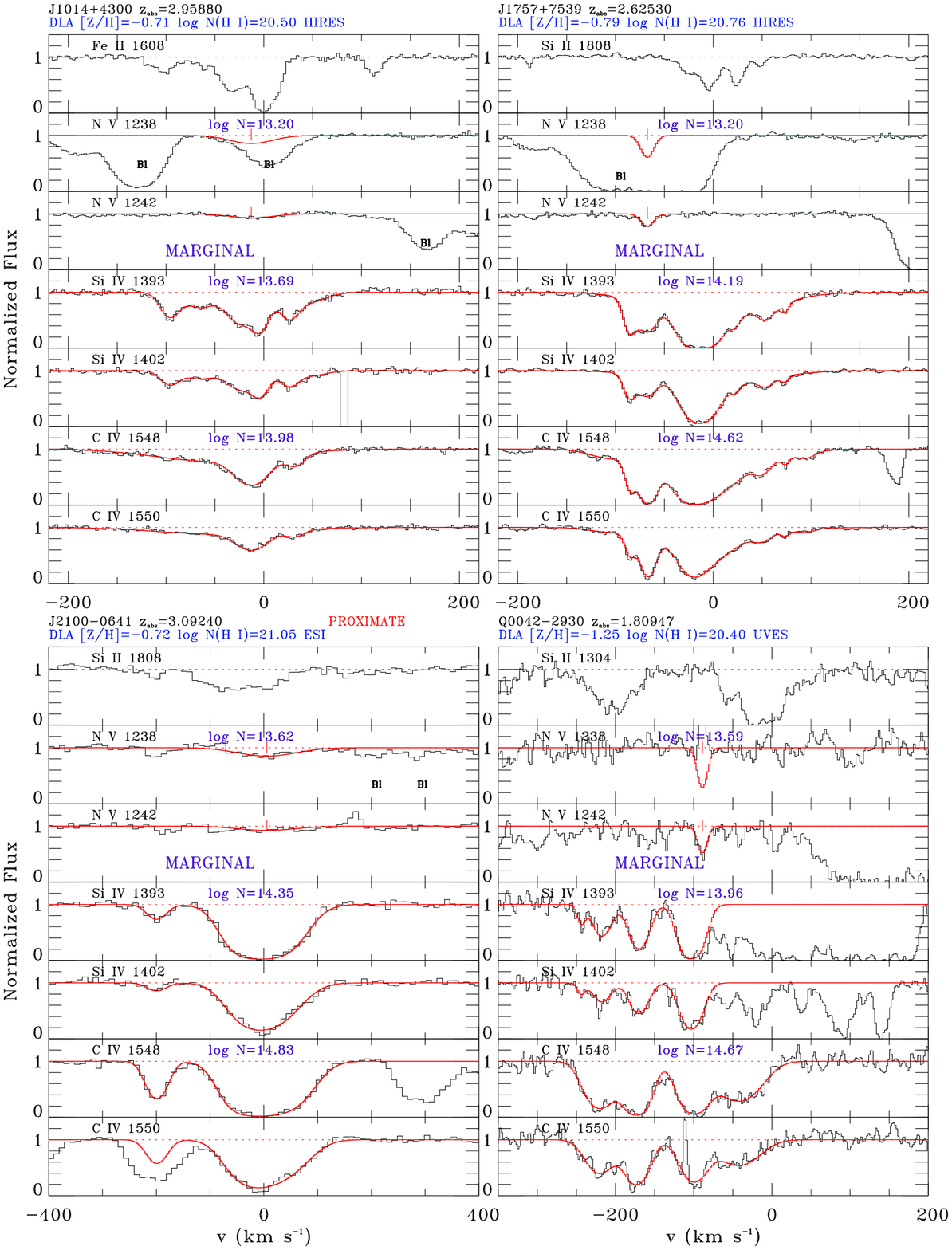}
\caption{High-ion absorption-line profiles for the four DLAs
  with marginal \nf\ detections. A low-ionization line is also shown.
  VPFIT Voigt-profile models are included as solid red lines.
  In the \nf\ panels, blends are indicated with the label `Bl', and
  tick marks show the component centers in our best-fit model.
  The velocity zero-point is defined by the DLA redshift.
  We annotate on the high-ion panels the total column density
  obtained by VPFIT.}
\end{figure*}
\begin{figure*}
\includegraphics[width=18.5cm]{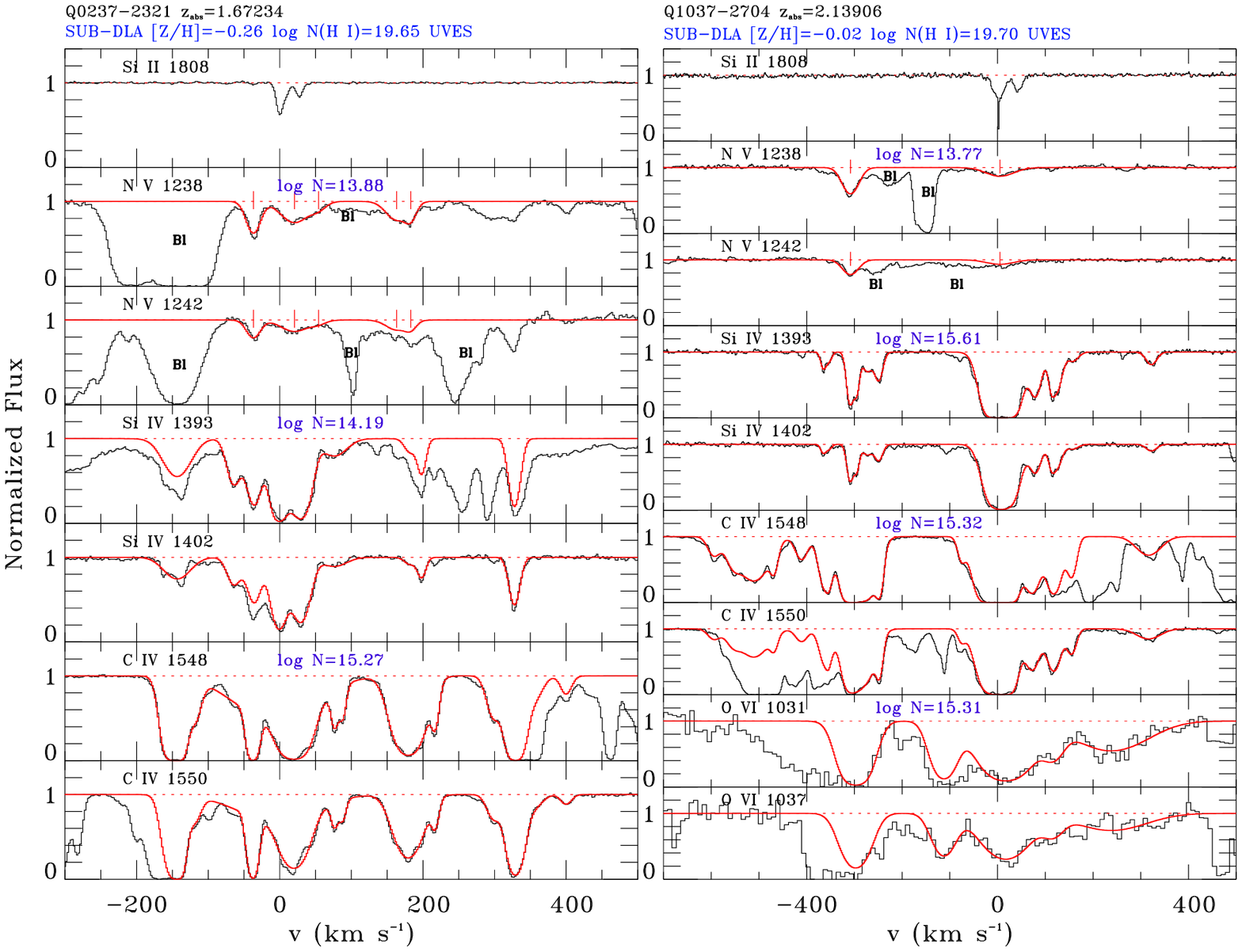}
\caption{High-ion absorption-line profiles for the two sub-DLAs
  with \nf\ detections. A low-ionization line is also shown.
  VPFIT Voigt-profile models are included as solid red lines.
  In the \nf\ panels, blends are indicated with the label `Bl', and
  tick marks show the component centers in our best-fit model.
  The velocity zero-point is defined by the DLA redshift.
  We annotate on the high-ion panels the total column density
  obtained by VPFIT.}
\end{figure*}

\begin{figure*}
\includegraphics[width=18.5cm]{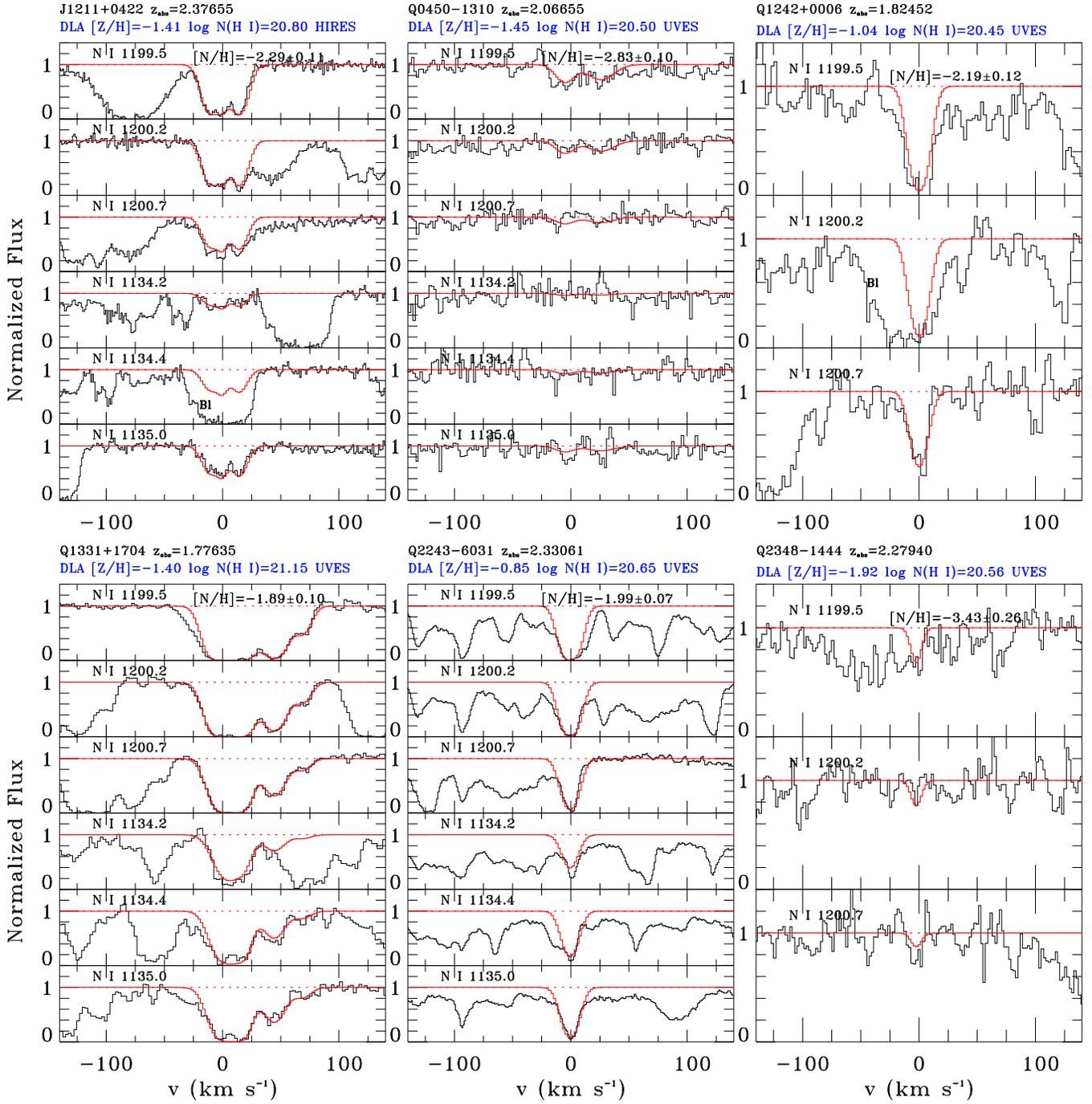}
\caption{Normalized \none\ absorption-line profiles in six DLAs with \nf.
  VPFIT Voigt component models are included as
  solid red lines. Blends are identified only when they appear at the
  velocities of \none\ absorption.}
\end{figure*}

\begin{figure}
\includegraphics[width=18.5cm]{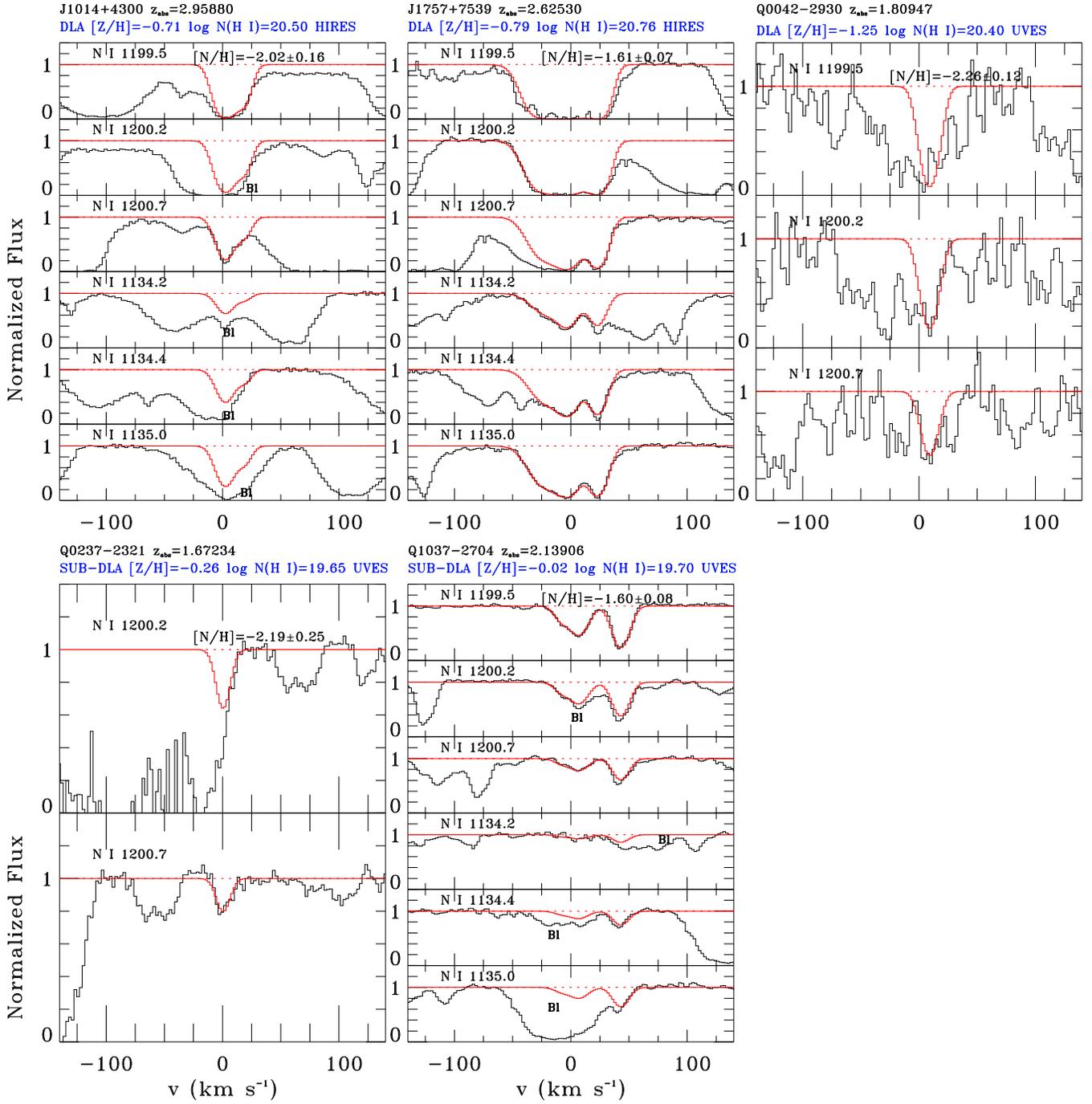}
\caption{Normalized \none\ absorption lines in three DLAs and two
  sub-DLAs with \nf. VPFIT Voigt component models are included as
  solid red lines. Blends are identified only when they appear at the
  velocities of \none\ absorption.} 
\end{figure}

\end{document}